%% file: paper.tex
\documentclass[runningheads]{llncs}
\usepackage[misc,geometry]{ifsym}
\usepackage[T1]{fontenc}

\usepackage{rotating}

\usepackage{amsfonts}
\usepackage{amsmath} 
\usepackage{xspace, mathtools, comment}
\usepackage[para,online,flushleft]{threeparttable}
\usepackage{booktabs, multirow, soul}
\usepackage{enumitem}
\usepackage{graphicx}
\usepackage{caption}
\usepackage{tabularx}
\usepackage{multicol}
\usepackage{xcolor}
\input{define}

\usepackage{floatrow}
\newlength\mysep 
\floatstyle{plaintop}
\restylefloat{table}

\usepackage{subfig}
\DeclareCaptionFormat{upper}{#1#2\uppercase{#3}\par}

\usepackage{algorithm,algpseudocode}

\algnewcommand{\LineComment}[1]{\Statex \(\triangleright\) #1}
\algnewcommand{\ShortLineComment}[1]{\Statex \hspace{1.8em}\(\triangleright\) #1}
\algnewcommand{\ShortShortLineComment}[1]{\Statex \hspace{3.1em}\(\triangleright\) #1}
\algnewcommand{\ShortShortShortLineComment}[1]{\Statex \hspace{4.4em}\(\triangleright\) #1}

\begin{document}
\title{T-Cell Receptor Optimization with Reinforcement Learning and Mutation Polices for Precision Immunotherapy}
\author{Ziqi Chen\inst{1}\and
Martin Renqiang Min\inst{2}$^{(\text{\Letter})}$ \and
Hongyu Guo \inst{3} \and
Chao Cheng \inst{4} \and
Trevor Clancy \inst{5} \and
Xia Ning \inst{1,6,7}$^{(\text{\Letter})}$}
\authorrunning{Chen et al.}
\titlerunning{T-Cell Receptor Optimization with Reinforcement Learning}
\institute{
Computer Science and Engineering, 
The Ohio State University, Columbus, OH 43210, USA \\
\email{\{ning.104\}@osu.edu} \and
Machine Learning Department, NEC Labs, Princeton, NJ 08540, USA \\
\email{\{renqiang\}@nec-labs.com} \and
Digital Technologies Research Centre, National Research Council Canada, Ontario, Canada \and
Department of Medicine, Baylor College of Medicine, Houston, TX 77030, USA \and
NEC Oncolmmunity AS, Oslo Cancer Cluster, Innovation Park, Ullernchauss\'{e}en 64, 0379, Oslo, Norway \and 
Biomedical Informatics, The Ohio State University, Columbus, OH 43210, USA \and 
Translational Data Analytics Institute, The Ohio State University, Columbus, OH 43210, USA \\
}

\maketitle              

\vspace{-10pt}
\begin{abstract}
T cells monitor the health status of cells by identifying foreign peptides displayed on their surface. 
T-cell receptors (TCRs), which are protein complexes found on the surface of T cells, are able to bind to these peptides. 
This process is known as TCR recognition and constitutes a key step for immune response. 
Optimizing TCR sequences for TCR recognition represents a fundamental step towards the development of personalized treatments to trigger immune responses killing cancerous or virus-infected cells. 
In this paper, we formulated the search for these optimized TCRs as a reinforcement learning (\RL) problem, and presented a framework 
\tcrppo with a mutation policy using proximal policy optimization. 
\tcrppo mutates TCRs into effective ones that can recognize given peptides.
{\tcrppo} leverages a reward function that combines the likelihoods of mutated sequences being valid TCRs measured by a new scoring function based on deep autoencoders, 
with the probabilities of mutated sequences recognizing peptides from a peptide-TCR interaction predictor.
We compared \tcrppo with multiple baseline methods and demonstrated that \tcrppo significantly outperforms all the baseline methods to generate 
positive binding and valid TCRs. These results demonstrate the potential of \tcrppo 
for both precision immunotherapy and peptide-recognizing TCR motif discovery.

\vspace{-2pt}
\end{abstract}

\vspace{-10pt}
\section{Introduction}
\vspace{-10pt}
%
Immunotherapy is a fundamental treatment for human diseases, which uses a person's immune system to fight diseases~\cite{Verdegaal2016,Esfahani2020,Waldman2020}. 
In the immune system, 
immune response is triggered by cytotoxic T cells {which are activated by the engagement of 
the T cell receptors (TCRs) with immunogenic peptides 
presented by Major Histocompatibility Complex (MHC) proteins on the surface of infected or cancerous cells.}
The recognition of these foreign peptides is determined by the interactions between the peptides and 
TCRs on the surface of T cells. 
This process is known as TCR recognition and constitutes a key step for immune response~\cite{Craiu1997,glanville2017identifying}. 
{Adoptive T cell immunotherapy (ACT), which has been a promising cancer treatment, genetically modifies the autologous T cells taken from patients in laboratory experiments, after which the modified T cells are infused into patients' bodies to fight cancer.}
As one type of ACT therapies, TCR T cell (TCR-T) therapy directly modifies the TCRs of T cells to increase the binding affinities, which makes it possible to recognize and kill tumor cells effectively~\cite{NatureSurvey2017}.
TCR is a heterodimeric protein with an $\alpha$ chain and a $\beta$ chain. 
Each chain has three loops as complementary determining regions (CDR): CDR1, CDR2 and CDR3. CDR1 and CDR2 are primarily responsible for interactions with MHC, and CDR3 interacts with peptides~\cite{rossjohn2015t}. 
The CDR3 of the $\beta$ chain has a higher degree of variations and is therefore arguably mainly responsible for the recognition of foreign peptides~\cite{la2018understanding}. 
In this paper, we focused on the optimization of the CDR3 sequence of $\beta$ chain in TCRs to enhance their binding affinities against 
peptide antigens, and we conducted the optimization through novel reinforcement learning. 
The success of our approach will have the potential to guide TCR-T therapy design. 
For the sake of simplicity, when we refer to TCRs in the rest of the paper, we mean the CDR3 of $\beta$ chain in TCRs.

Despite the significant promise of TCR-T therapy, optimizing TCRs for therapeutic purposes remains a time-consuming process, 
which typically requires exhaustive screening for high-affinity TCRs, either \emph{in vitro} or \emph{in silico}.
To accelerate this process, computational methods have been developed recently to predict peptide-TCR interactions~{\cite{Springer2020}}, 
leveraging the experimental peptide-TCR binding data~\cite{Shugay2017,Tickotsky2017} and TCR sequences~\cite{tcrdb2020}.
However, these peptide-TCR binding prediction tools cannot immediately direct the rational design of new high-affinity TCRs. 
Existing computational methods for biological sequence design include search-based methods~\cite{Arnold1998}, generative methods~\cite{Killoran2017,Gupta2019}, 
optimization-based methods~{\cite{Gonzalez2015}} and reinforcement learning ({\RL})-based methods~\cite{angerm2020,marcin2020}.
However, all these methods generate sequences without considering additional conditions such as peptides, 
and thus cannot optimize TCRs tailored to recognizing different
peptides.
In addition, these methods do not consider the validity of generated sequences, which is important for TCR optimization 
as valid TCRs should follow specific characteristics~{\cite{Hou2016}}.

In this paper, we presented a new reinforcement-learning ({\RL}) framework based on proximal policy optimization (PPO)~\cite{schulman17ppo}, referred to as {\tcrppo}~\footnote{The code is available at https://github.com/ninglab/TCRPPO}, to computationally optimize 
TCRs through a mutation policy. 
In particular, \tcrppo learns a joint policy to optimize TCRs customized for any given peptides.
In \tcrppo, we designed a new reward function that measures both the likelihoods of the mutated sequences being valid TCRs, 
and the probabilities of the TCRs recognizing peptides.  
To measure TCR validity, we developed a TCR auto-encoder, referred to as \autoenc, and utilized reconstruction errors from \autoenc
and also its latent space distributions, quantified by a Gaussian Mixture Model, to calculate novel validity scores. 
To measure peptide recognition, we leveraged a state-of-the-art peptide-TCR binding predictor \ergo~{\cite{Springer2020}}
to predict peptide-TCR binding. 
{Please note that \tcrppo is a flexible framework, as \ergo can be replaced by any other 
binding predictors~\cite{Cai2022,Weber2021}.}
In addition, we designed a novel buffering mechanism, 
referred to as {\buffer}, 
to revise TCRs that are difficult to optimize. 
We conducted extensive experiments using 
7 million TCRs from {\tcrdb}~{\cite{tcrdb2020}}, 10 peptides from {\mcpas}~{\cite{Tickotsky2017}} and 15 peptides from {\vdjdb}~{\cite{Shugay2017}}.
Our experimental results demonstrated that 
{\tcrppo} can substantially outperform the best baselines with best improvement of {45.04\% and 52.89\%} 
in terms of generating qualified TCRs with high validity scores and high recognition probabilities, over {\mcpas} and {\vdjdb} peptides, respectively.
Figure~\ref{fig:methods:architecture} presents the overall architecture of {\tcrppo}.

\vspace{-10pt}
\section{Related Work}
\vspace{-10pt}

Existing methods developed for biological sequence design include search-based methods, 
deep generative methods, optimization-based methods and \RL-based methods.
Among search-based methods, the classical evolutionary search~\cite{Arnold1998} uses an evolution strategy to 
randomly mutate the sequences and select desired ones in an iterative way.
Among generative methods, Killoran \etal~\cite{Killoran2017} 
optimized the latent embeddings of DNA sequences learned from a variational autoencoder towards better properties.
Gupta \etal~{\cite{Gupta2019}} used generative adversarial networks (GANs) to generate DNA sequences and selected the generated ones
with desired properties to further optimize GANs.
Among optimization-based methods, Gonzalez \etal~{\cite{Gonzalez2015}} used a Gaussian process model 
to emulate the production rates of a certain protein across different gene designs in living cells, and then optimized the 
gene designs to improve the production rates using Bayesian optimization.
Both the above generative methods and optimization methods aim at optimizing the biological sequences
without any additional conditions,
As a consequence, these methods are not applicable to our TCR optimization problem. 
In our problem, the optimization of TCRs must be tailored to given peptides, 
because TCRs binding to different peptides have different characteristics.

Despite the success of {\RL} on many applications, there remains limited work of 
applying {\RL} to biological sequence design.
Angermueller \etal~\cite{angerm2020} developed a model-based {\RL} method for biological sequence design using PPO to improve the sample efficiency, 
where the policy is trained over a simulator model learned to approximate the reward function.
Skwark \etal~\cite{marcin2020} then leveraged a {\RL} method based on PPO to discover a potential Covid-19 cure.
Their method aims at identifying the variants of human angiotensin-converting enzyme (ACE2) protein sequence that have higher binding affinities 
against the SARS-CoV-2 spike protein than the original ACE2 protein. 
Our \tcrppo also applies PPO with a mutation policy to optimize TCR sequences.
However, \tcrppo is fundamentally different from previous methods in three aspects.
First, {\tcrppo} learns a joint policy to optimize the TCRs customized for any given peptides.
In addition, {\tcrppo} employs a comprehensive reward function to simultaneously optimize the validity and 
the recognition probability of the TCRs against the peptides.
{\tcrppo} also leverages a buffering mechanism to generalize the optimization capability of {\tcrppo} to TCRs that are hard to optimize, 
or peptides with fewer positive binding TCRs.

\vspace{-10pt}
\section{Methods}
\vspace{-5pt}

\vspace{-20pt}
\begin{minipage}{\textwidth}
	\vspace{-15pt}
	\begin{minipage}{0.5\textwidth}
		\begin{figure}[H]
			\includegraphics[width=.90\linewidth]{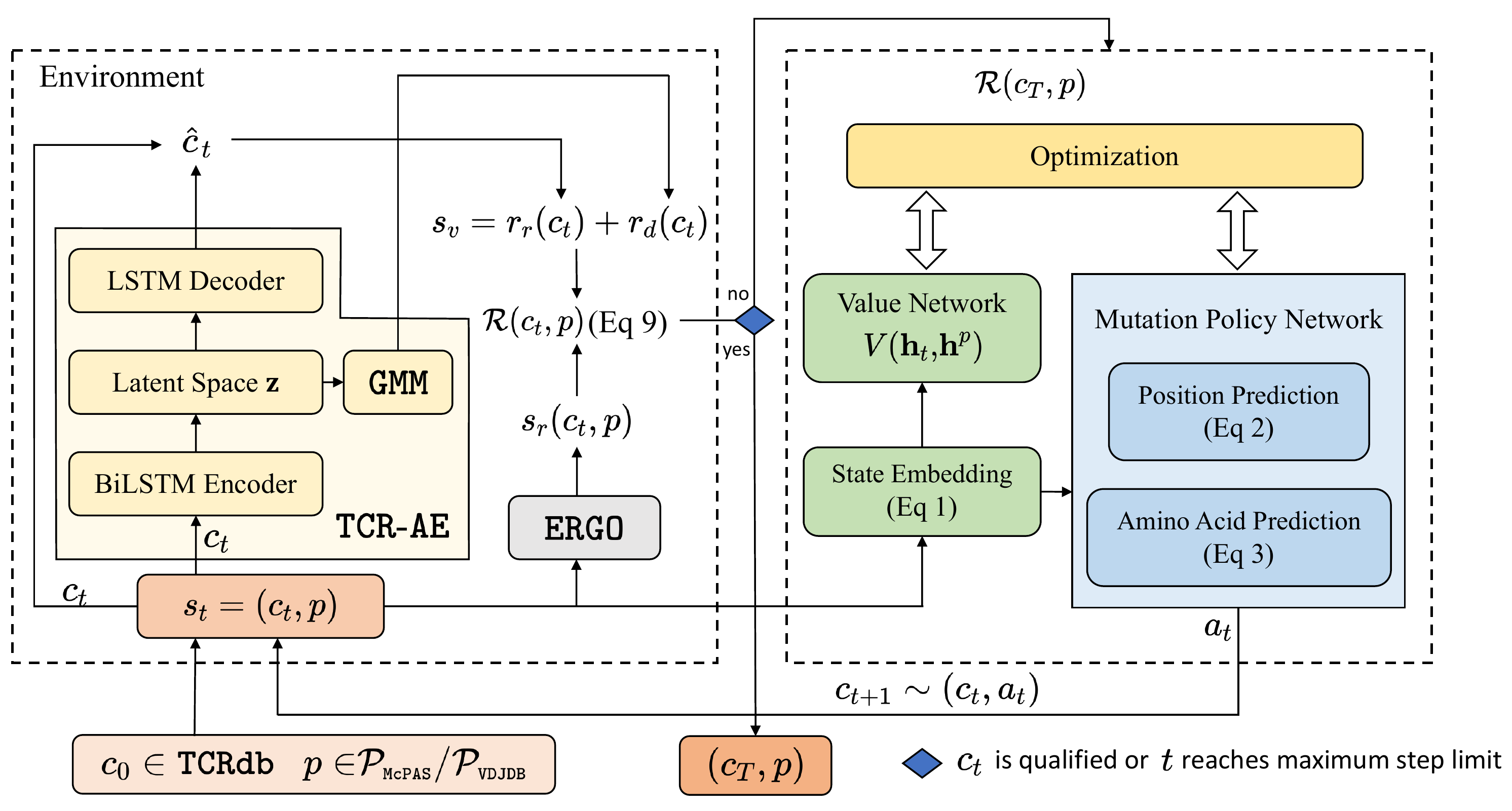}
			\caption{{Model Architecture of {\tcrppo}}}
			\label{fig:methods:architecture}
		\end{figure}
	\end{minipage}
	\hfill
	\begin{minipage}{0.43\textwidth}
		\begin{figure}[H]
			\vspace{14pt}
			\includegraphics[width=1.05\linewidth]{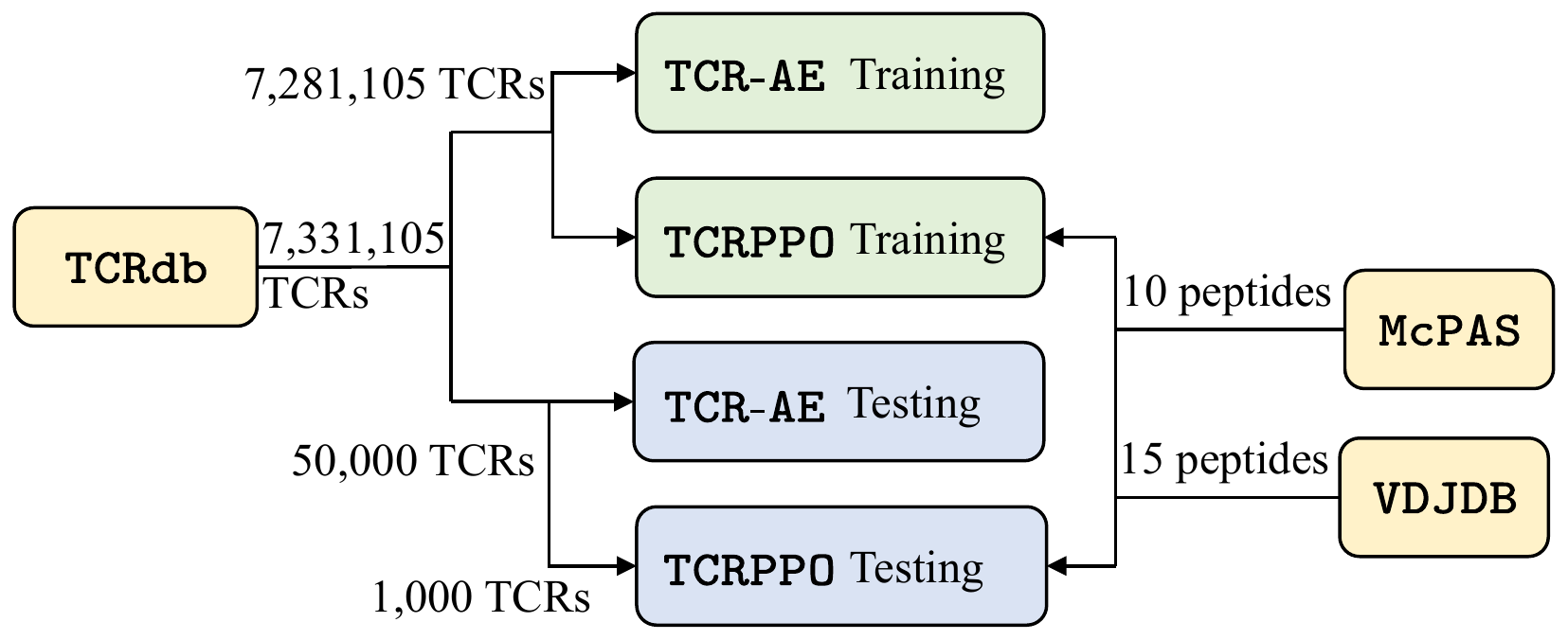}
			\vspace{6pt}
			\caption{Data Flow for {\tcrppo} and {\autoenc} training and testing}
			\label{fig:statistics}
		\end{figure}
	\end{minipage}
	\vspace{-25pt}
\end{minipage}

\subsection{Problem Definition}
\label{sec:method:problem}
%

In this paper, the recognition ability of a TCR sequence against the given peptides is measured by a recognition probability, denoted as {\recog}.
The likelihood of a sequence being a valid TCR is measured by a score, denoted as {\istcr}.
A \emph{qualified} TCR is defined as a sequence with $\recog>\sigma_{\text{r}}$ and {$\istcr>\sigma_c$}, 
where $\sigma_r$ and {$\sigma_{c}$} 
are pre-defined thresholds {($\sigma_r$=0.9 and $\sigma_c$=1.2577, as discussed in Appendix~\ref{appendix:setup} and 
{\ref{appendix:result:reward}}, respectively)}.
The goal of \tcrppo is to mutate the existing TCR sequences that have low recognition probability against the given peptide, into 
qualified ones.
A peptide \peptide or a TCR sequence \tcr is represented as a sequence of its amino acids
{$\langle\amino_1, \amino_2, \cdots, \amino_i, \cdots, \amino_l\rangle$}, 
where $\amino_i$ is one of the 20 types of natural amino acids at the position $i$ in the sequence, 
and $l$ is the sequence length.
We formulated the TCR mutation process as a Markov Decision Process (MDP) $M=\{\statespace,\actionspace,P,\mathcal{R}\}$
containing the following components: 

\begin{itemize}[leftmargin=*,noitemsep]

\item \statespace: the state space, in which each state $\sta\in\statespace$ is a tuple of 
a potential TCR sequence \tcr and a peptide \peptide, that is, $\sta = (\tcr, \peptide)$. Subscript $t$ ($t=0, \cdots, T$) is used to index step of 
\sta, that is, $\sta_t = (\tcr_t, \peptide)$.
Please note that $\tcr_t$ may not be a valid TCR. 
A state $\sta_t$ is a terminal state, denoted as $\sta_T$, if it contains a qualified $\tcr_t$, or $t$ reaches the maximum step limit $T$.
Please also note that \peptide will be sampled at $\sta_0$ and will not change over time $t$,
\item \actionspace: the action space, in which each action $\action\in\actionspace$ is a tuple of a mutation site $i$ and a mutant amino acid $\amino$, that is, $\action= (i, \amino)$. 
Thus, the action will mutate the amino acid at position $i$ of a sequence \mbox{\tcr = $\langle\amino_1, \amino_2, \cdots, \amino_i, \cdots, \amino_l\rangle$} into another amino acid \amino. 
Note that \amino has to be different from $\amino_i$ in \tcr.
\item \transit: the state transition probabilities, in which $\transit(\sta_{t+1}|\sta_{t},\action_t)$ specifies 
	the probability of next state $\sta_{t+1}$ at time $t+1$ from state $\sta_{t}$ at time $t$
	with the action $\action_t$. 
	In our problem, the transition to $\sta_{t+1}$ is deterministic, that is $\transit(\sta_{t+1}|\sta_{t},\action_t)=1$.
\item \reward: the reward function at a state. In \tcrppo, all the intermediate rewards at states $\sta_t$ ($t=0, \cdots, T-1$)
are 0; only the final reward at $\sta_T$ is used to guide the optimization. 

\end{itemize}

\vspace{-20pt}
\subsection{Mutation Policy Network}
\label{sec:method:policy}
\vspace{-5pt}
%
%
\tcrppo mutates one amino acid in a sequence \tcr at a step to modify \tcr into a qualified TCR. 
Specifically, at the initial step $t=0$, a peptide \peptide is sampled as the target, and a valid TCR 
$\tcr_0$ is sampled to initialize $\sta_0=(\tcr_0, \peptide)$; 
at a state $\sta_t = (\tcr_t, \peptide)$ ($t>0$), the mutation policy network of \tcrppo predicts an action $\action_t$
that mutates one amino acid of $\tcr_t$ to modify it into $\tcr_{t+1}$ that is more likely to lead to a final, qualified TCR bound to \peptide. 
\tcrppo encodes the TCRs and peptides in a distributed embedding space.
It then learns
a mapping between the embedding space and the mutation policy, as discussed below.

\vspace{-10pt}
\subsubsection{Encoding of Amino Acids}
\label{sec:method:encode}

Following the idea in Chen \etal~\cite{Chen2021}, 
we represented each amino acid \amino by concatenating three vectors: 1) $\aminoemb^b$, the corresponding 
row of \amino in 
the BLOSUM matrix, 2) $\aminoemb^o$, the one-hot encoding of \amino, and 3) $\aminoemb^d$, the learnable embedding, 
that is, \amino is encoded as \mbox{$\aminoemb = \aminoemb^b \oplus \aminoemb^o \oplus \aminoemb^d$},
where $\oplus$ represents the concatenation operation.
We used such a mixture of encoding methods to enrich the representations of amino acids within \tcr and \peptide.
%

\vspace{-10pt}
\subsubsection{Embedding of States}
\label{sec:method:embed} 

We embedded $\sta_t = (\tcr_t, \peptide)$ via embedding its associated sequences $\tcr_t$ and \peptide. 
For each amino acid $\amino_{i,t}$ in $\tcr_t$, 
we embedded $\amino_{i,t}$ and its context information in $\tcr_t$ into a hidden vector $\hidden_{i,t}$
using a one-layer bidirectional LSTM~\cite{Graves2005} as below,
\begin{equation}
\label{eqn:lstm}
\begin{aligned}
\overrightarrow{\hidden}_{i,t}, \overrightarrow{\cell}_{i,t}\!& =\!\!\text{LSTM}(\aminoemb_{i,t}, \overrightarrow{\hidden}_{i-1,t}, \overrightarrow{\cell}_{i-1,t}; \overrightarrow{W}); \\
\overleftarrow{\hidden}_{i,t}, \overleftarrow{\cell}_{i,t}\!& =\!\!\text{LSTM}(\aminoemb_{i,t}, \overleftarrow{\hidden}_{i+1,t}, \overleftarrow{\cell}_{i+1,t}; \overleftarrow{W}); \\
\hidden_{i,t}\!\!& =\!\!\overrightarrow{\hidden}_{i,t} \oplus \overleftarrow{\hidden}_{i,t}
\end{aligned}
\end{equation}
where $\overrightarrow{\hidden}_{i,t}$ and $\overleftarrow{\hidden}_{i,t}$ are the hidden state vectors of the $i$-th amino acid in $\tcr_t$; 
$\overrightarrow{\cell}_{i,t}$ and $\overleftarrow{\cell}_{i,t}$ are the memory cell states of $i$-th amino acid;
$\overrightarrow{W}$ and $\overleftarrow{W}$ are the learnable parameters of the two LSTM directions, respectively; and 
$\overrightarrow{\hidden}_{0,t}$,$\overleftarrow{\hidden}_{l_c,t}$, $\overrightarrow{\cell}_{0,t}$ and $\overleftarrow{\cell}_{l_c,t}$ 
($l_c$ is the length of $\tcr_t$) are initialized with random vectors. 
With the embeddings of all the amino acids, we defined the embedding of $\tcr_t$ as the concatenation of hidden vectors at the two ends, that is, 
$\hidden_t = \overrightarrow{\hidden}_{l_c,t} \oplus \overleftarrow{\hidden}_{0,t}$.
We embedded a peptide sequence into a hidden vector $\hidden^{\scriptsize{\peptide}}$ using another bidirectional LSTM in the same way.
%

\vspace{-10pt}
\subsubsection{Action Prediction}
%

To predict the action $\action_t = (i, \amino)$ at time $t$, \tcrppo needs to make two predictions: 1) the position $i$ of current 
$\tcr_t$ where $\action_t$ needs to occur; 2) the new amino acid $\amino$ that $\action_t$ needs to place with at position $i$. 
To measure ``how likely" the position $i$ in $\tcr_t$ is the action site, \tcrppo uses the following network: 
\vspace{-5pt}
\begin{equation}
\label{eqn:position}
f(i) = \vect{w}^\mathsf{T}(\text{ReLU}(W_1\hidden_{i, t} + W_2\hidden^{\scriptsize{\peptide}})) / (\sum\nolimits_{j=1}^{l_c}\vect{w}^\mathsf{T}(\text{ReLU}(W_1\hidden_{j, t} + W_2\hidden^{\scriptsize{\peptide}}))),
\end{equation}
where $\hidden_{i,t}$ is the latent vector of $\amino_{i,t}$ in $\tcr_t$ (Equation~\ref{eqn:lstm});
$\hidden^{\scriptsize{\peptide}}$ is the latent vector of \peptide; 
$\vect{w}$/$W_j$ ($j$=1,2) are the learnable vector/matrices.
Thus, \tcrppo measures the probability of position $i$ being the action site by 
looking at its context encoded in $\hidden_{i,t}$ and the peptide \peptide.
The predicted position $i$ is sampled from the probability distribution from Equation~\ref{eqn:position}
to ensure necessary exploration. 

Given the predicted position $i$, \tcrppo needs to predict the new amino acid that should replace $\amino_i$ in $\tcr_t$. 
\tcrppo calculates the probability of each amino acid type being the new replacement as follows: 
\begin{equation}
\vspace{-2pt}
g(\amino) = \text{softmax}(U_1 \times \text{ReLU}(U_2\hidden_{i,t}+U_3\hidden^{\scriptsize{\peptide}})), 
\label{eqn:amino}
\end{equation}  
where $U_j$ ($j$=1,2,3) are the learnable matrices;
$\text{softmax}(\cdot)$ converts a 20-dimensional vector into probabilities over the 20 amino acid types.
The replacement amino acid type is then determined by sampling from the distribution, 
excluding the original type of $\amino_{i,t}$.

\vspace{-10pt}
\subsection{Potential TCR Validity Measurement}
\label{sec:method:istcr}

Leveraging the literature~\cite{zong2018deep,Abati2019}, 
we designed a novel scoring function to quantitatively measure the likelihood of a given sequence \tcr being a valid TCR
(i.e., to calculate \istcr), which will be part of the reward of \tcrppo.
Specifically, we trained a novel auto-encoder 
model, denoted as \autoenc, from only valid TCRs.
We used the reconstruction accuracy of a 
sequence in \autoenc to measure its TCR validity. 
The intuition is that since \autoenc is trained from only valid TCRs, its encoding-decoding process will obey the ``rules" of 
true TCR sequences, and thus, a non-TCR sequence could not be well reproduced from \autoenc. 
However, it is still possible that a non-TCR sequence can receive a high reconstruction accuracy from \autoenc, 
if \autoenc learns some generic patterns shared by TCRs and non-TCRs and fails to detect irregularities,
or \autoenc has high model complexity~\cite{Pang2021,zong2018deep}.
To mitigate this, we additionally evaluated the latent space within \autoenc using a Gaussian Mixture Model (\gmm), hypothesizing that 
non-TCRs would deviate from the dense regions of TCRs in the latent space.

\vspace{-10pt}
\subsubsection{\autoenc}	

Figure~\ref{fig:methods:architecture} presents the auto-encoder \autoenc. 
\autoenc uses a bidirectional LSTM to encode an input sequence \tcr into $\hidden^\prime$ by concatenating the last hidden vectors from the two LSTM directions (similarly as in Equation~\ref{eqn:lstm}). 
{Please note that this bidirectional LSTM is independent of the mutation policy network in \tcrppo.}
$\hidden^\prime$ is then mapped into a latent embedding $\latent^\prime$ as follows, 
%
\vspace{-5pt}
\begin{equation}
\label{eqn:z}
\latent^\prime = W^z\hidden^\prime,
\vspace{-5pt}
\end{equation} 
which will be decoded back to a sequence $\hat{\tcr}$ via a decoder. 
The decoder has a single-directional LSTM that decodes $\latent^\prime$ by generating one amino acid at a time as follows, 
\vspace{-5pt}
\begin{equation}
\hidden^\prime_{i}, \cell^\prime_{i}  = \text{LSTM}(\hat{\aminoemb}_{{i-1}}, \hidden^\prime_{i-1}, \cell^\prime_{i-1}; W^\prime); \ \ 
\hat{o}_i = \text{softmax}(U^\prime \times \text{ReLU}(U^\prime_{1}\hidden^\prime_{i}+U^\prime_{2}\latent^\prime)), 
\vspace{-5pt}
\end{equation}
where $\hat{\aminoemb}_{{i-1}}$ 
is the encoding of the amino acid $\hat{o}_{i-1}$ that is decoded from step $i-1$; 
$W^\prime$ is the parameter. 
The LSTM starts with a zero vector $\aminoemb_0 = \mathbf{0}$ and $\hidden_0 = W^h\latent^\prime$.
The decoder infers the next amino acid 
by looking at the previously decoded amino acids encoded in $\hidden^\prime_{i}$ and 
the entire prospective sequence encoded in $\latent^\prime$.

Please note that \autoenc is trained from TCRs, independently of \tcrppo and in an end-to-end fashion. 
Teacher forcing~{\cite{Williams1989}} is applied during training to {feed the ground truth amino acids as inputs to predict the next amino acid, and thus cross entropy loss is applied on each amino acid to optimize \autoenc.}
As a stand-alone module, \autoenc is used to calculate the score {\istcr. 
The input sequence \tcr to \autoenc is encoded using only BLOSUM matrix as 
we found empirically that BLOSUM encoding can lead to a good reconstruction performance and a fast convergence compared to 
other combinations of encoding methods. 

\vspace{-10pt}
\subsubsection{Reconstruction-based score}

With a well-trained \autoenc, we calculated the reconstruction-based TCR validity score of a sequence \tcr as follows, 
\vspace{-5pt}
\begin{equation}
\reconst(\tcr) = 1 - {\mathtt{lev}(\tcr, \autoenc(\tcr))}/{l_{\scriptsize{\tcr}}}
\vspace{-5pt}
\end{equation}
where $\autoenc(\tcr)$ represents the reconstructed sequence of {\tcr} from {\autoenc}; 
$\mathtt{lev}(.)$ is the Levenshtein distance, an edit-distance-based metric, between
 \tcr and $\autoenc(\tcr)$; $l_{\scriptsize{\tcr}}$ is the length of \tcr.
Higher $\reconst(\tcr)$ indicates higher probability of \tcr being a valid TCR.
Please note that when \autoenc is used in testing, the length of the reconstructed sequence might not be the same 
as the input \tcr, because
\autoenc could fail to accurately predict the end of the sequence, leading to either too short or too long reconstructed sequences.
Therefore, we normalized the Levenshtein distance using the length of input sequence $l_{\scriptsize{\tcr}}$ similarly to Snover {\etal}~{\cite{Snover2006}}.
Please note that {$\reconst(\tcr)$} could be negative when the distance is greater than the sequence length.
The negative values will not affect the use of the scores (i.e., negative $\reconst(\tcr)$ indicates very different 
$\autoenc(\tcr)$ and \tcr).

\vspace{-5pt}
\subsubsection{Density Estimation-based Score}

To better distinguish valid TCRs from invalid ones, 
\tcrppo also conducts a density estimation over the latent space of $\latent^\prime$ (Equation~\ref{eqn:z})
using \gmm.  For a given sequence \tcr, \tcrppo calculates the likelihood score of \tcr falling within 
the Gaussian mixture region of training TCRs as follows, 
\vspace{-5pt}
\begin{equation}
\dense(\tcr) = \exp(1 + \frac{\log P(\latent^{\prime})}{\tau})
\vspace{-5pt}
\label{eqn:reward:density}
\end{equation}
where $\log P(\latent^{\prime})$ is the log-likelihood of the latent embedding 
$\latent^{\prime}$; $\tau$ is a 
constant used to rescale the log-likelihood value ($\tau = 10$).
We carefully selected the parameter $\tau$ such that 90\% of TCRs can have $\dense(\tcr)$ above $0.5$.
As we do not have invalid TCRs, we cannot use classification-based scaling methods such as Platt scaling~{\cite{Platt1999}} to 
calibrate the log likelihood values to probabilities.
%

\vspace{-10pt}
\subsubsection{TCR Validity Scoring}

Combining the reconstruction-based scoring and density estimation-based scoring, we developed a new 
scoring method 
to measure TCR validity as follows:
\vspace{-5pt}
\begin{equation}
\label{eqn:istcr}
\istcr(\tcr) = r_r(\tcr) + r_d(\tcr). 
\vspace{-5pt}
\end{equation}
This method is used to evaluate if a sequence is likely to be a valid TCR and is used in the reward function.

\vspace{-15pt}
\subsection{\tcrppo Learning}

\vspace{-7pt}
\subsubsection{Final Reward}

We defined the final reward for \tcrppo based on \recog and \istcr scores as follows, 
\vspace{-5pt}
\begin{equation}
\label{eqn:reward}
\reward(\tcr_T, \peptide) = \recog(\tcr_T, \peptide) + \alpha \min(0, \istcr(\tcr_T)-\istcrsigma)
\vspace{-5pt}
\end{equation}
where $\recog(\tcr_T, \peptide)$ is the predicted recognition probability by \ergo, 
\istcrsigma is a threshold that $\tcr_T$ is very likely to be a valid TCR; 
and $\alpha$ is the hyperparameter used to control the tradeoff between \recog and \istcr ($\alpha=0.5$).
%

\vspace{-15pt}
\subsubsection{Policy Learning}
\label{sec:method:optimization}

We adopted the proximal policy optimization (PPO)~\cite{schulman17ppo} 
to optimize the policy network as discussed in Section~\ref{sec:method:policy}. 
The objective function of PPO is defined as follows:
\begin{equation}
\vspace{-7pt}
\begin{aligned}
\max\nolimits_{\boldsymbol{\Theta}} L^{\text{CLIP}}(\boldsymbol{\Theta}) & = \hat{\mathbb{E}}_t[\min(r_t(\boldsymbol{\Theta})\hat{A}_t, \text{clip}(r_t(\boldsymbol{\Theta}), 1-\epsilon, 1+\epsilon)\hat{A}_t)], \\
& \ \text{where}\  
r_t(\boldsymbol{\Theta}) = \frac{\pi_{{\boldsymbol{\Theta}}}(a_t|s_t)}{\pi_{{\boldsymbol{\Theta}_{\text{old}}}}(a_t|s_t)}, 
\end{aligned}
\end{equation}
where $\boldsymbol{\Theta}$ is the set of learnable parameters of the policy network and $r_t(\boldsymbol{\Theta})$ is the probability ratio 
between the action under current policy $\pi_{{{\boldsymbol{\Theta}}}}$ and the action under previous policy $\pi_{{\boldsymbol{\Theta}_{\text{old}}}}$.
Here, $r_t(\boldsymbol{\Theta})$ is
clipped to avoid moving $r_t$ outside of the interval $[1-\epsilon, 1+\epsilon]$.
$\hat{A}_t$ is the advantage at timestep $t$ computed with the generalized advantage estimator~\cite{Schulmanetal2016}, measuring how much better a selected action is than 
others on average:
\vspace{-3pt}
\begin{equation}
\hat{A}_t = \delta_t + (\gamma\lambda)\delta_{t+1} + ... + (\gamma\lambda)^{T-t+1}\delta_{T-1},
\end{equation}
where $\gamma\in(0, 1)$ is the discount factor determining the importance of future rewards; 
$\delta_t = r_t + \gamma V(s_{t+1}) - V(s_t)$ is the temporal difference error in which $V(s_t)$ is a value function;
$\lambda\in(0,1)$ is a parameter used to balance the bias 
and variance of $V(s_t)$.
Here, $V(\cdot)$ uses a multi-layer perceptron (MLP) to predict the future return of current state $s_t$ from the peptide embedding $\hidden^{\scriptsize{\peptide}}$ and the TCR embedding $\hidden_t$.
The objective function of $V(\cdot)$ is  as follows: 
\vspace{-5pt}
\begin{equation}
\min\nolimits_{\boldsymbol{\Theta}} L^{V}(\boldsymbol{\Theta}) = \hat{\mathbb{E}}_t[(V(\hidden_t, \hidden^{\scriptsize{\peptide}}) - \hat{R}_t)^2],
\label{eqn:value}
\end{equation}
where $\hat{R}_t = \sum_{i=t+1}^T \gamma^{i-t} r_i$ is the rewards-to-go.
Because we only used the final rewards, that is $r_i = 0 \text{ if } i \neq T$, we calculated $\hat{R}_t$ with $\hat{R}_t = \gamma^{T-t} r_{\scriptsize{T}}$.
We also added the entropy regularization loss $H(\boldsymbol{\Theta})$, 
a popular strategy used for policy gradient methods~\cite{mniha16,schulman17ppo}, to encourage the exploration of the policy.
The final objective function of \tcrppo is defined as below,
\vspace{-5pt}
\begin{equation}
\min\nolimits_{\boldsymbol{\Theta}} L(\boldsymbol{\Theta}) =  - L^{\text{CLIP}}(\boldsymbol{\Theta}) + \alpha_1L^{V}(\boldsymbol{\Theta}) - \alpha_2H(\boldsymbol{\Theta}), 
\vspace{-5pt}
\end{equation}
where {$\alpha_1$} and {$\alpha_2$} are two hyperparameters controlling the tradeoff among the PPO objective, the value function and the entropy regularization term.

\vspace{-10pt}
\subsubsection{Reward-Informed Buffering and Re-optimization}
\label{sec:method:buffer}
%
\tcrppo implements a novel buffering and re-optimizing mechanism, denoted as \buffer, 
to deal with TCRs that are difficult to optimize, 
and to generalize its optimization capacity to more, diverse TCRs. 
{To optimize TCRs, various number of mutations will be applied to get the binding TCRs.
For TCRs requiring more mutations, it could be more difficult for \tcrppo to optimize; and thus re-optimizing these TCRs enables \tcrppo to explore more actions for the optimization of difficult TCRs, instead of being overwhelmed by relatively simple cases.}
This mechanism includes a buffer, which memorizes the TCRs that cannot be optimized to qualify. 
These hard sequences {and the corresponding peptides} will be sampled from the buffer again following the probability distribution below, to be further optimized by 
\tcrppo, 
\begin{equation}
\label{eqn:S}
\vspace{-5pt}
S(\tcr,\peptide) = \xi^{(1-\scriptsize{\reward(\tcr_T, \peptide)})}/\Sigma.
\end{equation}
In Equation~\ref{eqn:S}, $S$ measures how difficult to optimize \tcr against \peptide based on its final reward $\reward(\tcr_T, \peptide)$ 
in the previous optimization, 
$\xi$ is hyper-parameter ($\xi=5$ in our experiments), and $\Sigma$ converts $S(\tcr,\peptide)$ as a probability. 
It is expected that by doing the sampling and re-optimization, \tcrppo is better trained to learn from hard sequences, and also the 
hard sequences have the opportunity to be better optimized by \tcrppo. 
In case a hard sequence still cannot be optimized to qualify, it will 
have 50\% chance of being allocated back to the buffer. 
In case the buffer is full (size 2,000 in our experiments), the sequences earliest allocated in the buffer 
will be removed. 
We referred to the \tcrppo with \buffer as \tcrppobuf.

\vspace{-10pt}
\section{Experimental Settings}
\label{sec:experiments}

\vspace{-10pt}
\subsection{Datasets}
\label{sec:settings:dataset}
\vspace{-5pt}

We selected peptides and TCR sequences for the training and testing of {\tcrppo} and {\autoenc}.
Figure~\ref{fig:statistics} summaries the peptides and TCRs used in our experiments.

\vspace{-10pt}
\subsubsection{Peptides}
To test \tcrppo, we first identified a set of peptides that \tcrppo needs to optimize TCR sequences for. We aimed at selecting the 
peptides which {are very likely to }have reliable peptide-TCR binding predictions, 
such that their binding predictions can serve to test \tcrppo's optimized TCRs against the respective peptides. 
We identified such peptides from two databases: \mcpas~\cite{Tickotsky2017} and \vdjdb~\cite{Shugay2017},
which have experimentally validated TCR-peptide binding pairs. 
We applied the autoencoder-based \ergo models~\cite{Springer2020} 
on \mcpas and \vdjdb (\mcpas and \vdjdb have pre-specified training and testing sets for each peptide), 
and selected the peptides which have AUC values above 0.9 on their respective testing sets.  
This resulted in 10 peptides selected from \mcpas, denoted as \mcpasP, and 15 peptides selected from \vdjdb, denoted as \vdjdbP, 
with lengths ranging from 8 to 21{, as presented in Appendix Table~\ref{tbl:peptides}}. 
Additional discussion on peptides is available in Appendix~\ref{appendix:data}.

%
\vspace{-10pt}
\subsubsection{TCR sequences}
%
We then selected TCR sequences that \tcrppo needs to optimize against each of the peptides selected as above. 
We selected such sequences from the \tcrdb database~\cite{tcrdb2020}, which contains  
277 million human TCR sequences, each with a \mbox{TCR-$\beta$} sequence. 
Here, we used \tcrdb, not \mcpas or \vdjdb, because \tcrdb's sequences are valid TCRs and have no information on 
their binding affinities with the selected peptides. 
Therefore, these valid TCRs can be used to train \autoenc to calculate \istcr. 
Meanwhile, since \tcrdb is much larger than \mcpas (4,528 TCRs) and \vdjdb (50,049 TCRs), it is very likely that 
the \istcr calculated from the \autoenc, which is trained over \tcrdb data, 
will be independent of the \recog calculated from \ergo, 
which is trained on \mcpas or \vdjdb, avoiding possible correlation between \istcr and \recog as in the reward (Equation~\ref{eqn:reward}).  

We selected all the TCRs with lengths below 27 (\ergo can only predict sequences of length 27 or shorter) from \tcrdb, 
resulting in 7,331,105 unique TCR-$\beta$ sequences. 
{Figure~\ref{fig:length_dist} presents the distribution of lengths of TCRs in \tcrdb.
As shown in Figure~\ref{fig:length_dist}, the most common length of TCRs is 15.}
Additional discussion on the length of TCRs is available in Appendix~\ref{appendix:data}.
Among these selected sequences, we sampled 50K sequences, denoted 
as the validation set \Validation, to test and validate \istcr; 
within these 50K sequences, we again sampled 1K sequences, denoted as the testing set \Test, to test \tcrppo performance once it is well trained. 
The remaining selected sequences (i.e., not in the validation set), denoted as the training set \Train, 
are used to train \tcrppo; 
they are also used to train \autoenc. 

\vspace{-15pt}
\subsection{Experimental Setup}
\label{sec:settings:setup}
\vspace{-5pt}

For all the selected peptides from a same database (i.e., 10 peptides from \mcpas, 15 peptides from \vdjdb), we trained one \tcrppo agent, which 
optimizes the training sequences (i.e., 7,281,105 TCRs in Figure~\ref{fig:statistics}) to be qualified against one of the selected peptides. 
The \ergo model trained on the corresponding database (the same {\ergo} model also used to select the peptides from the database 
as in Section~\ref{sec:settings:dataset}) will be used to test recognition probabilities \recog for the \tcrppo agent. 
Please note that as in Springer \etal~{\cite{Springer2020}}, one {\ergo} model is trained for all the peptides in each database 
(i.e., one {\ergo} predicts TCR-peptide binding for multiple peptides). Thus, the \ergo model is suitable to test \recog for multiple peptides in our setting.
Also note that we trained one \tcrppo agent corresponding to each database, because peptides and TCRs in these two databases 
are very different, demonstrated by the inferior performance of an \ergo trained over the two databases together, and discussed in Springer 
\etal~{\cite{Springer2020}}.

\tcrppo mutates each sequence up to 8 steps (i.e., T=8).
In \tcrppo training, an initial TCR sequence (i.e., $\tcr_0$ in $\sta_0$) is randomly sampled from \Train, 
and will be mutated in the following states; 
a peptide \peptide is randomly sampled at $\sta_0$, and remains the same in the following states (i.e., $\sta_t = (\tcr_t, \peptide)$).  
Once a \tcrppo is well trained from \Train, it will be tested on \Test. 
{We set the dimensions of the hidden layers (e.g., hidden layers of action prediction networks) as 256, and 
the dimensions of latent embeddings (e.g., $\hidden^{\scriptsize{\peptide}}$, $\hidden_t$) as 128 (i.e., half of the hidden dimensions).}
Other hyper-parameters {and the details of hyper-parameter selection} of the TCR mutation environment, the policy network and the \RL 
agent are available in Appendix~\ref{appendix:setup}.

\input{tables/result_full_new}

\vspace{-15pt}
\subsection{Baseline Methods}
\vspace{-5pt}

We compared the \tcrppo method and \tcrppobuf with multiple baseline methods of two primary categories: 
1) generative methods that generate a new TCR in its entirety, and 
2) mutation-based methods that optimize TCRs via mutating amino acids of existing TCRs. 
For generative methods, we used two baseline methods including Monte Carlo tree search ({\mcts})~\cite{Coulom2007} and
a variational autoencoder with backpropagation ({\bpvae})~\cite{GomezBombarelli2018}. 
For mutation-based methods, we used three baseline methods to mutate each TCR sequence up to 8 steps and stop the mutation once a qualified TCR is generated, 
including random mutation ({\randommutate}), greedy mutation ({\greedy}) and genetic mutation ({\genetic})~\cite{Whitley1994}. 
In addition, we used a random selection method (\randomselect) as another baseline to randomly sample a TCR from \tcrdb, which helps quantify the space of valid TCRs. 
More details about baseline methods are available in Appendix~{\ref{appendix:baseline}}.

\vspace{-10pt}
\subsection{Evaluation Metrics}
\vspace{-5pt}
We evaluated all the methods using six metrics including:
(1) qualification rate {\qr}, which measures the percentage of qualified TCRs (Section~\ref{sec:method:problem}) among all the output TCRs;
(2) edit distance between $\tcr_0$ and $\tcr_T$ ({\mutediff}), which measures sequence difference between 
$\tcr_0$ and qualified $\tcr_T$;
(3) average TCR validity score {\avgistcr} over valid TCRs  or over qualified TCRs; 
(4) average recognition probability {\avgrecog} over valid TCRs or over qualified TCRs; 
(5) validity rate {\vr}, which measures the percentage of valid TCRs (Section~\ref{sec:method:problem}) among all the output TCRs;
(6) average number of calls to {\reward} calculation ({\ncalls}) (Equation~\ref{eqn:reward}) over all the generated TCRs, which estimates 
the efficiency of methods. 
We calculated the metrics over two different sets of output TCRs:
(1) the set of valid TCRs {\vTCRs}: $\vTCRs = \{\tcr| \istcr(\tcr) > \istcrsigma\}$;
(2) the set of qualified TCRs {\qTCRs}: ${\qTCRs} = \{\tcr| \recog(\tcr)>\istcrsigma, \tcr\in\vTCRs\}$.

%

\vspace{-10pt}
\section{Experimental Results}
\label{sec:experiment_result}

\vspace{-10pt}
\subsection{Comparison on TCR Optimization Methods}
\label{sec:study:mutation}
\vspace{-5pt}

\subsubsection{Overall Comparison}

Table~\ref{tbl:perform:all} presents the overall comparison among all the methods over  \mcpasP and \vdjdbP. 
In \mcpasP, {\tcrppo} methods achieve overall the best performance: 
in terms of \qr, \tcrppo achieves {\mbox{$25.89\!\pm\!29.60\%$}},  
which {slightly outperforms} the best \qr from the baseline method \genetic (\mbox{$25.18\!\pm\!21.28\%$}). 
\tcrppobuf achieves the best \qr at {\mbox{$36.52\!\pm\!30.25\%$}} 
among all the methods, which is {45.04\%} 
better than the best from the baseline methods ($25.18\!\pm\!21.28\%$ from \genetic). 
\tcrppobuf achieves so  with a few \reward calls (\ncalls=7).
In {\vdjdbP}, {\tcrppo} methods also achieve overall the best performance: in terms of {\qr}, {\tcrppo} and {\tcrppobuf} outperform the best 
baseline \genetic by {18.80\% and 52.89\%.} 

Among the qualified TCRs (\qTCRs) for \mcpasP, 
\tcrppo {and \tcrppobuf} methods achieve the highest $\avgistcr$ values on average ({$1.55\!\pm\!0.19$} 
for \tcrppo; 
{$1.55\!\pm\!0.17$} 
for \tcrppobuf), 
with above {6}\% improvement 
from those of \genetic, which achieves the best \qr among all the baseline methods. 
Note that qualified TCRs must have \istcr above \istcrsigma, which is set as 1.2577 as discussed 
in Appendix~\ref{appendix:result:reward}. The significant high \istcr values from \tcrppo methods demonstrate that \tcrppo is able to generate qualified TCRs that are highly likely to be valid TCRs. 
In terms of \recog among qualified TCRs, \tcrppo methods have $\avgrecog(\qTCRs)$ value 
{$0.97\!\pm\!0.03$}, above the \prsigma. 
Note that $\prsigma=0.9$, 
a very high threshold for \recog to determine TCR-peptide binding, is actually a very tough constraint. The fact that \tcrppo methods can survive this constraint with substantially 
high \qr and highly likely valid TCRs as results demonstrates the strong capability of \tcrppo methods. 
In addition, \tcrppo methods need a few number of calls to calculate \reward (i.e., {7} 
for \tcrppo, compared to 166 for \genetic), indicating that 
\tcrppo is very efficient in identifying qualified TCRs.
In \vdjdbP, we observed very similar trends as those in \mcpasP.

Among all the baseline methods, 
\randomselect randomly selects a valid TCR from \tcrdb, given that some TCRs in \tcrdb may already qualify. Thus, it is a naive baseline for all the other methods. 
It has \vr around 95\%, corresponding to how our \istcr threshold \istcrsigma is identified (by 95 percentile true positive rate) as discussed in Appendix~\ref{appendix:result:reward}. 
On average, among all valid TCRs, about 0.05\% (\qr in \randomselect) TCRs are qualified TCRs. \randommutate, \greedy, \genetic and \tcrppo methods substantially 
outperform this baseline method. 

\vspace{-12pt}
\subsubsection{Comparison among Mutation-based Methods}

\randommutate, \greedy, \genetic, \tcrppo and \tcrppobuf are mutation-based methods: they start from a valid TCR from \tcrdb, 
and optimize the TCR by mutating its amino acids. 
Among all the mutation-based methods, \tcrppo and \tcrppobuf outperform others as discussed above. 
Below we only focused on \mcpasP, as similar trends exist on \vdjdbP.

In \mcpasP, 
\randommutate underperforms all the other mutation-based methods: it has \qr below 3\% on average, but has very high \vr (close to 100\%). 
\randommutate uses \reward to select randomly mutated sequences, which decomposes to its \istcr and its \recog components. 
\randommutate starts from valid TCRs with high \istcr already, but low \recog in general. 
During random mutation, \recog cannot be easily improved as no knowledge or informed guidance is used to direct the mutation towards 
better \recog. 
Therefore, the \istcr component in \reward will dominate, leading to that the final selected, best mutated sequence tends to 
have high \istcr to satisfy high \reward. 
Such best mutated sequence tends to occur after only a few random mutations, since as shown in Appendix~\ref{appendix:result:reward}, 
random mutations can quickly decrease \istcr values. 
Thus, the qualified sequences produced from \randommutate tend to be more similar to the initial TCRs (\mutediff is small, around 3; 
high \vr as $99.36\!\pm\!0.21\%$), 
their \istcr values are high (close to 1.5; highest among all mutation-based methods) but \qr is low due to hardly improved \recog values.

In \mcpasP, 
\greedy has a better \qr ($20.58\!\pm\!17.85\%$) than \randommutate and also a high \vr ($99.98\!\pm\!0.04\%$). 
\greedy leverages a greedy strategy to select the random mutation that gives the best \reward at the current step. 
Thus, it leverages some guidance on \recog improvement based on \reward, and can improve \recog 
values compared to \randommutate. As in Table~\ref{tbl:perform:all}, it has better $\recog(\vTCRs)$ value ($0.62\!\pm\!0.12$) for valid TCRs. 
Meanwhile, \greedy explores a large sequence space, allowing its to identify more valid TCRs, leading to high \vr ($99.98\!\pm\!0.04$),
high \qr ($20.58\!\pm\!17.85$; also due to better \recog improvement) but more diverse results ($\mutediff =5.10\!\pm\!1.09$) than \randommutate.

\genetic is the second best mutation-based method on \mcpasP, after \tcrppo methods. 
\genetic explores a sequence space even larger than that in \greedy, and uses \reward to guide next mutations also in a greedy way. 
Thus, it enjoys the opportunity to reach more, potentially qualified TCRs, demonstrated by high \mutediff (\mbox{$5.16\!\pm\!1.05$}) indicating diverse sequences, and thus
achieves a better \qr ($25.18\!\pm\!21.28\%$) than \greedy, even better than that of \tcrppo, and a high \vr ($100.00\pm0.00\%$), 
at a significant cost of many more calls to calculate \reward. 
Even though, it still significantly underperforms \tcrppobuf in terms of \qr and 
qualities of qualified TCRs as discussed earlier. 

\vspace{-10pt}
\subsubsection{Comparison between Mutation-based Methods and Generation-based Methods}

Overall, mutation-based methods substantially outperform generation-based methods. For example, in terms of \qr,
mutation-based methods (excluding random mutation \randommutate) has an average \qr $22.88\%$ in \mcpasP, 
compared to $0.03\%$ of the generation-based methods (\mcts and \bpvae). 
In addition to the superior performance, mutation-based methods have strong biological relevance 
that make them very suitable and practical for TCR-T based precision immunotherapy, 
in which they can be readily employed to optimize existing TCRs found in patient's TCR repertoire.  
However, 
any promising TCRs generated from generation-based methods have to be either synthesized, which could be  
both very costly and technically challenging, or mutated from existing TCRs that are similar to the generated TCRs in their 
amino acids. 
Detailed discussions on generation-based methods are available in Appendix~\ref{appendix:result:discuss}.

\vspace{-10pt}
\subsection{Evaluation on Optimized TCR Sequences}
\label{sec:experiment_result:tcr}

\begin{figure}[H]
	\vspace{-15pt}
	\centering
	\sidesubfloat[]{%
		\includegraphics[width=0.5\textwidth]{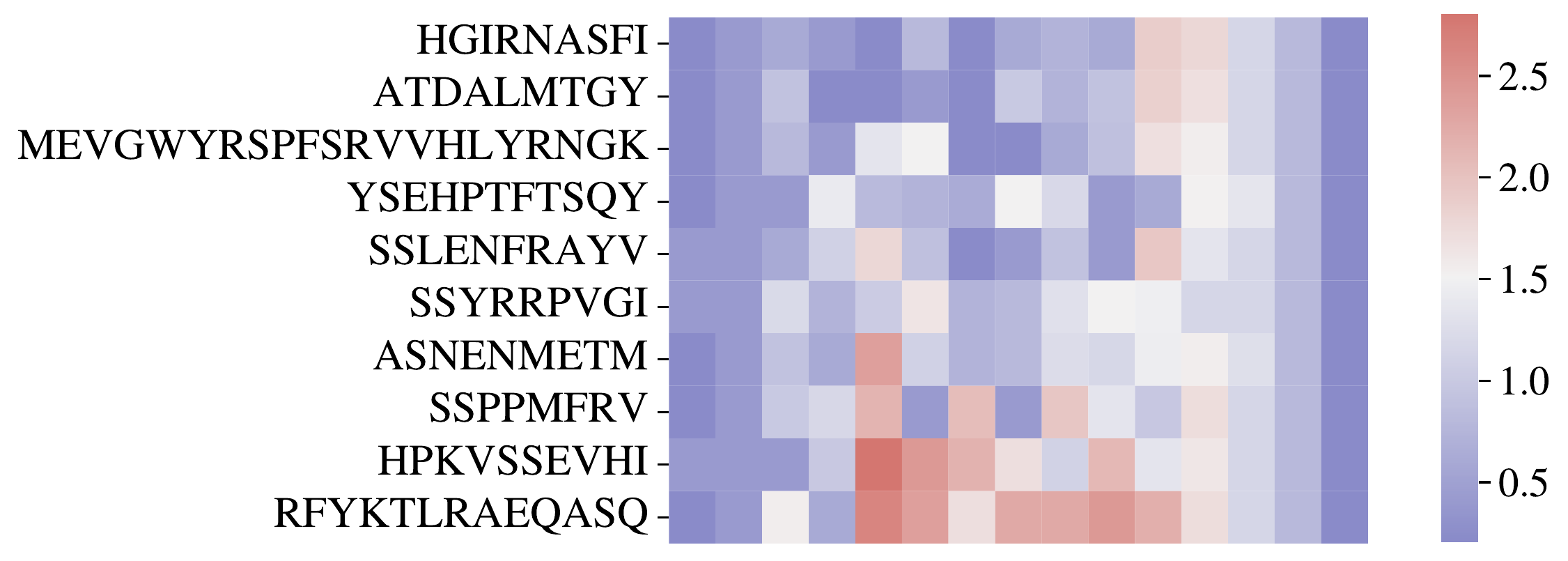}%
		\label{fig:figure3A}
	}
	\hspace{0.01\textwidth}
	\sidesubfloat[]{%
		\includegraphics[width=0.3\textwidth]{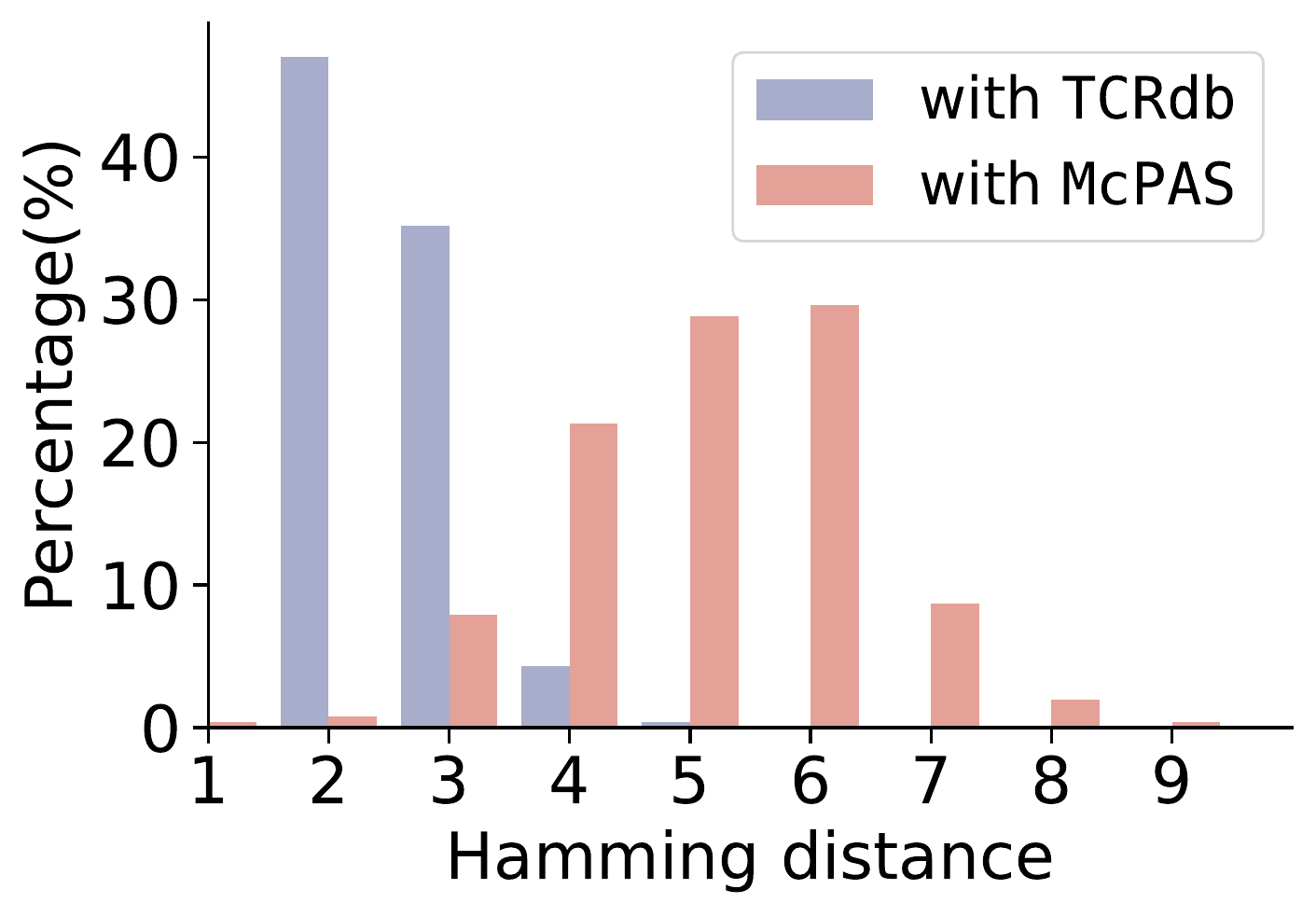}%
		\label{fig:figure3B}
	}
	\vspace{-12pt}
	\caption{Optimized TCR patterns for \mcpas peptides (\textbf{A}); {TCR distances (\textbf{B})}.}
	\label{fig:figure3}
\end{figure}

%

\vspace{-15pt}
Figure~\ref{fig:figure3}A presents the entropy of amino acid distributions at each sequence position 
among the length-15 TCRs for each \mcpasP peptide. Here the TCRs are optimized by \tcrppo with respect to the \mcpas 
peptides.
Figure~\ref{fig:figure3}A clearly shows some common patterns among all the optimized TCRs. For example, 
the first three positions and last four positions tend to have high conservation. TCRs for some peptides (e.g.,  
``SSPPMFRV", ``ASNENMETM", and ``RFYKTLRAEQASQ") have high variations at internal regions. 
Similar patterns are also observed among the binding TCRs for other peptides in \vdjdb~\footnote{https://vdjdb.cdr3.net/motif}.

Figure~\ref{fig:figure3}B presents the difference between qualified TCRs optimized by \tcrppo and existing TCRs.
It presents the distribution of Hamming distances between qualified TCRs 
(length 15) for peptide {``RFYKTLRAEQASQ''} 
and their most similar (in terms of Hamming distance) 
TCRs that are known to bind to this peptide in \mcpas;
and the distribution of Hamming 
distances between qualified TCRs for this peptide
and their most similar TCRs in \tcrdb.  
This figure demonstrates that the qualified TCRs by \tcrppo are actually different from known binding TCRs, 
but there are TCRs similar to them existing in \tcrdb. 
{This can be meaningful for precision immunotherapy,
as {\tcrppo} can produce diverse TCR candidates that are different from known TCRs,}
leading to a novel sequence space; meanwhile, 
these TCR candidates actually have similar human TCRs available for further medical
evaluation and investigation purposes.

Additional analyses are available in Appendix~\ref{appendix:result}, such as the overview of amino acid distributions of TCRs, the patterns for binding TCRs
and the comparison on TCR detection.
Specifically, we found that \tcrppo can successfully learn the patterns of TCRs (Appendix~\ref{appendix:result:overview}); we also found that \tcrppo can identify the specific binding patterns which are 
more conserved than the real binding patterns (Appendix~\ref{appendix:result:binding}).
We also found that our \istcr scoring method successfully distinguishes TCRs from non-TCRs in Appendix~\ref{appendix:result:reward}.

\vspace{-10pt}
\section{Conclusions {and Outlook}}
\label{sec:conclusion}
\vspace{-5pt}

In this paper, we presented a reinforcement learning framework to optimize TCRs for more effective TCR recognition, 
which has the potential to guide TCR engineering therapy.
Our experimental results in comparison with generation-based methods and mutation-based methods on optimizing TCRs demonstrate 
that \tcrppo outperforms the baseline methods.
Our analysis on the TCRs generated by \tcrppo demonstrates that \tcrppo can successfully learn the conservation patterns of TCRs.
Our experiments on the comparison between the generated TCRs and existing TCRs demonstrate that \tcrppo can generate TCRs similar to 
existing human TCRs, which can be used for further medical evaluation and investigation.
Our results in TCR detection comparison show that the \istcr score in our framework can very effectively detect non-TCR sequences. 
Our analysis on the distribution of \istcr scores over mutations demonstrates that \tcrppo mutates sequences along the trajectories not 
far away from valid TCRs.

{Our proposed \tcrppo is a modular and flexible framework. 
Thus, 
the \autoenc and \ergo scoring functions in the reward design can be replaced with other predictors trained on large-scale data when available. 
Also, \tcrppo can be further improved from the following perspectives. 
First, the recognition probabilities of TCRs considered in our paper 
are based on a peptide-TCR binding predictor (i.e., \ergo) rather than 
experimental validation.
Therefore, testing the generated qualified TCR candidates in a 
wet-lab will be needed ultimately to validate the interactions between TCRs and peptides.
Moreover, when the predicted recognition probabilities are not sufficiently accurate, 
\tcrppo learned from unreliable rewards could be inaccurate, resulting in 
generating 
TCRs that could not recognize the given peptide. 
Thus, it could be an interesting and challenging future 
 work 
to incorporate the reliabilities of predictions in the reward function of \tcrppo, so that the effect of unreliable recognition probabilities can be alleviated.
%
Finally, 
\tcrppo only considers the CDR3 of {$\beta$} chain in TCRs, while other regions of TCRs, though not contributing most to interactions between TCRs and peptides, are not considered.
In this sense, incorporating other regions of TCRs (e.g., CDR3 of alpha chains) could be an interesting future work.}

\bibliographystyle{splncs04}
\bibliography{bib}

\newpage

\setcounter{section}{0}
\renewcommand{\thesection}{A.\arabic{section}}

\setcounter{table}{0}
\renewcommand{\thetable}{A.\arabic{table}}

\setcounter{figure}{0}
\renewcommand{\thefigure}{A.\arabic{figure}}

\setcounter{algorithm}{0}
\renewcommand{\thealgorithm}{A.\arabic{algorithm}}

\setcounter{equation}{0}
\renewcommand{\theequation}{A.\arabic{equation}}

\vspace{10pt}

\section{{Data Analyses}}
\label{appendix:data}

{We listed the selected peptides from {\mcpas} and {\vdjdb} in Table~{\ref{tbl:peptides}}. 
As shown in Table~{\ref{tbl:peptides}}, these selected peptides are very different.
For example, the average edit distance between pairs of peptides in \mcpasP and \vdjdbP are 10.84 and 9.14, respectively, demonstrating the diversity of peptides compared with the average length of peptides in \mcpasP and \vdjdbP (10.90 and 9.67).
Comparing \mcpasP with \vdjdbP, we found that 6 peptides included in \mcpasP are also included in \vdjdbP, which could be due to the data overlapping between \mcpas and \vdjdb~\cite{Shugay2017}.
}

{We also listed the distribution of length of TCRs in \tcrdb in Figure~\ref{fig:length_dist}.
We found that the most common length of TCRs is 15; and TCRs with 13, 14, 15, and 16 lengths account for over 10\% of the entire database.
In this manuscript, our analyses focus on TCRs of length 13, 14, 15 and 16.
}

\input{tables/peptides}

\begin{figure}[!h] 
	\includegraphics[width=0.5\textwidth]{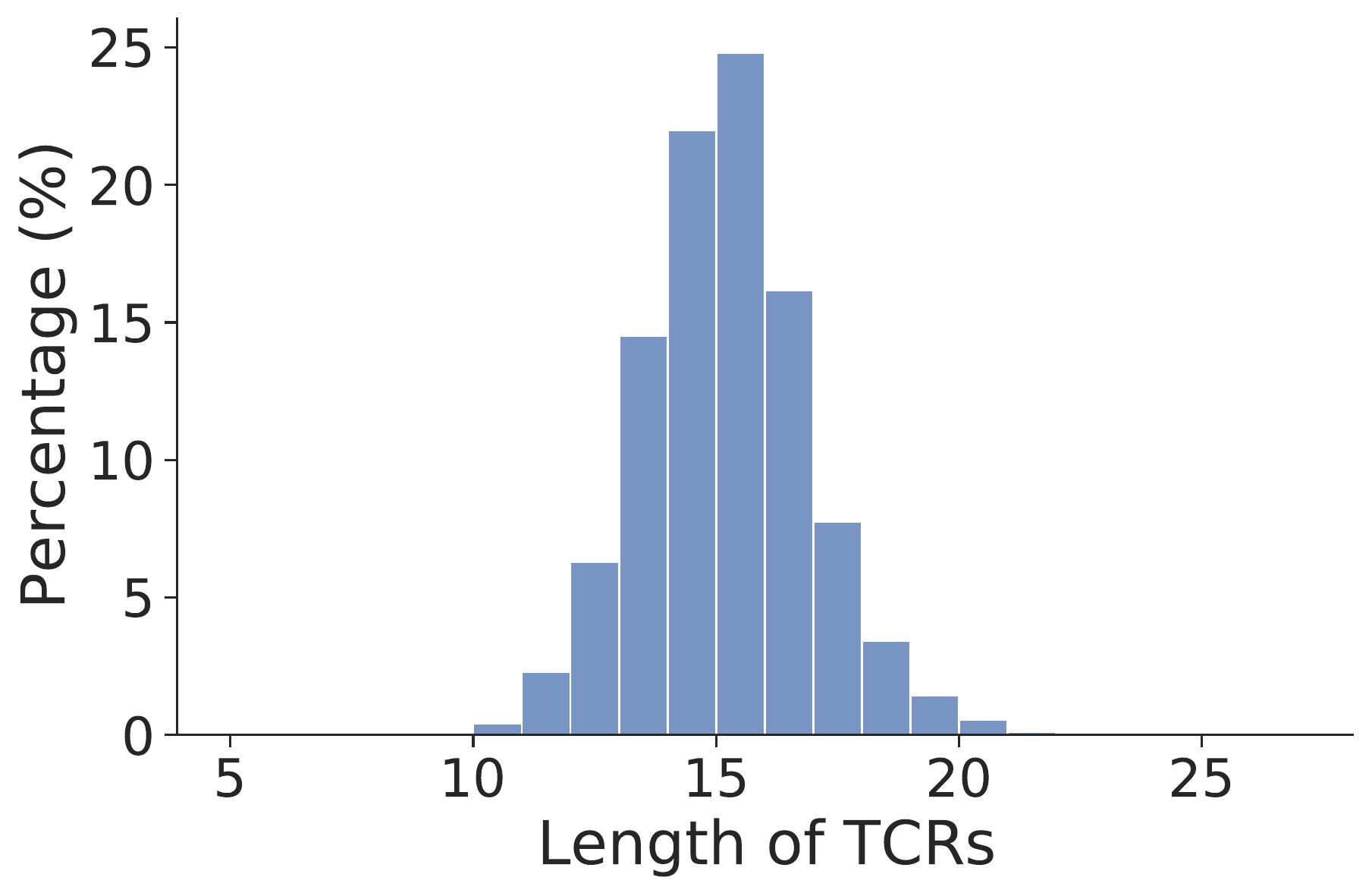}
	\caption{{The distribution of length of TCRs in {\tcrdb}.}}
	\label{fig:length_dist}
\end{figure}

\section{Parameter setup}
\label{appendix:setup}

\input{tables/setup}

We listed the hyper-parameters of the TCR mutation environment, the policy network and the \RL 
agent in Table~\ref{tbl:setup}.
We selected an optimal set of hyperparameters including the discount factor, 
the entropy coefficient and the maximum steps by a grid search,
according to the average rewards of the final 10 iterations.
{
Particularly, we set the maximum steps $T$ as 8 steps, as we empirically found that it can lead to the 
highest average final rewards.
We set the parameter $\alpha$ as 0.5 for the tradeoff between validity scores and recognition probabilities, as we empirically found that different $\alpha$ values could lead to comparable results.
We also constructed a validation set by sampling 1,000 TCRs from the test set of \autoenc;
these TCRs are not included in the test set of \tcrppo.
We then selected the hyperparameters including the hidden dimension of policy network and value networks
and the ratio of difficult initial state from \buffer that the corresponding models achieve
the maximum qualified percentage \qr.
Among three options $[64,128,256]$, the optimal hidden dimensions of policy and value networks for \mcpas and \vdjdb are 128. 
We set the latent dimension of $\overrightarrow{\hidden}_i/\overleftarrow{\hidden}_i/\hidden^p$ to be double of the optimal hidden dimensions, that is, 256.
In terms of encodings of amino acids, we set the dimension of learnable embeddings as 20, which is the same with the dimensions of BLOSUM embeddings and one-hot encodings; that is, the total dimension of amino acid encodings is 60.
In terms of ratios of difficult initial state, the optimal ratios for \mcpas and \vdjdb are 0.2 and 0.1, respectively.}
%
%
{We set the threshold for recognition probability \recog as 0.9, which is demonstrated as a proper threshold as it can lead 
to 47.04\% and 56.48\% at true positive rate and 0.16\% and 0.30\% at false positive rate, respectively, for selected peptides in \mcpas and \vdjdb.}
In the implementation of \tcrppo, for each iteration, we ran 20 environments in parallel for 256 timesteps and collect trajectories with 5,120 timesteps in total.
We trained \tcrppo for 1e7 timesteps (i.e., 1954 iterations). 
We set all the other hyper-parameters for the \RL agent as the default hyper-parameters provided by the stable-baselines3~\cite{stable-baselines3}, 
as we found little performance improvement from tuning these parameters.
%
%
%
{We trained the models using a Tesla P100 GPU and a CPU on 
Red Hat Enterprise 7.7. 
It took 9-10 hours to train a \tcrppo model.
With the trained model available, on a single GPU and a single CPU core, optimizing 1,000 TCRs for a single peptide using \tcrppo takes 160.5 seconds, which is faster than that of the state-of-the-art baseline \genetic (i.e., 207.8s).}

For \autoenc, we selected an optimal set of hyperparameters including the dimension of hidden layers and the dimension of latent layers. 
According to the reconstruction accuracy on the validation set, we set the dimension of hidden layers in \autoenc to be 64, and the dimension of 
latent layers to be 16.
At each step, we randomly sampled 256 TCR sequences from \tcrdb, and trained the \autoenc for 100,000 steps.
This \autoenc is with high reconstruction accuracy on our validation set, which is $94.6$\%, and thus can be used to provide reliable reconstruction-based scores for TCR 
sequences.

\section{Baseline methods}
\label{appendix:baseline}

We describe the baseline methods used in the main text as follows,
\begin{itemize}[leftmargin=*]
	\item \mcts: Monte Carlo Tree Search.
	Given a peptide \peptide, \mcts generates TCR sequences by adding one amino acid step by step
	until reaching the maximum length of 20 amino acids or with the recognition probability \recog greater than $\sigma_r$.
	All the generated peptides with no less than 10 amino acids will be evaluated by \ergo.
	The approach of \mcts is described in Algorithm~\ref{alg:mcts}.
        {At each step, the amino acid to be added is determined by the Upper Confidence Bound algorithm (see equation in line 12 of Algorithm~\ref{alg:mcts}) to balance exploration and exploitation~\cite{auer2002}.}
	The hyper-parameter $c_{puct}$ is set as 0.5.
	{For each peptide, the number of rollout $N$ is set as 1,000, as all the methods including \tcrppo generate 1,000 TCRs for each peptide.
        At the beginning of search, we initialize the sequence with ``C'', as all the TCRs in \tcrdb begin with ``C''.
	We collected all the generated TCRs during the search process, and 
        calculated the metrics listed in Table~\ref{tbl:perform:all} over the 1,000
	TCRs with the highest reward values for each peptide.}
	\item \bpvae: Back Propagation with the Variational Autoencoder.
	In \bpvae, a VAE model is first pre-trained to convert TCR sequences of variable lengths into continuous latent 
	embeddings of a fixed size with a single-layer LSTM.
	Then, a student model with the pre-trained VAE is employed to {
        distill knowledge from \ergo via learning the recognition probabilities produced by \ergo from the latent embeddings}~{{\cite{hinton2015}}}.
	Note that during this fine-tuning process, the parameters of pre-trained VAE will also be updated.
	With the well-trained student model, \bpvae can generate TCRs by optimizing the latent embeddings of TCRs through 
	gradient ascent to maximize the predicted recognition probabilities from the student model.
	The decoded TCR sequences with the prediction from student model greater than $\sigma_r$ will be evaluated by \reward function.
	We fine-tuned the hyper-parameters and set the dimension of hidden layers for VAE as 64, the dimension of latent embeddings as 8, 
	the batch size as 256, the {initial} coefficient for KL loss as 0.05 {and the initial learning rate as 0.005}.
        {We increased the coefficient by 0.05 every 1e4 steps until reaching the maximum coefficient limit 0.3; we also decreased the learning rate by 10 percent every 5,000 steps until reaching the minimum learning rate limit 0.0001.}
	We pre-trained the VAE model for 2e6 steps.
	We then fine-tune the VAE model with the student model for 50,000 steps.
	{During the inference, w}e set the maximum step $t$ for gradient ascent as 50 and the learning rate of gradient ascent as 0.05.
        {Similar to \mcts, we collected all the generated TCRs and calculated the metrics over the 1,000 TCRs with the highest reward values for each peptides.}
	\item \randommutate: random mutation.
	It randomly mutates a sequence one amino acid at a step up to 8 steps, each step leading to an intermediate mutated sequence; 
	it does so for each sequence $n$ times, 
	generating 8$n$ mutated sequences. It selects the mutated sequence with the highest \reward as the final output.  
	\item \greedy: greedy mutation. 
	It randomly mutates one amino acid of $\tcr_t$ at step $t$, and does this 10 times, generating 10 mutated sequences which are all 
	one-amino-acid different from $\tcr_t$. It selects the one with best \reward among the 10 sequences as the 
	next sequence to further mutate. 
	\item \genetic: genetic mutation.  
	It randomly mutates each sequence in the population (size $n=5$)  
	five times, each at one site, generating five mutated sequences. 
	Among all the mutated sequences 
	for all the sequences in the population, \genetic selects the top-$n$ sequences with the best $\reward$ as the next 
	population.
\end{itemize}

\begin{algorithm}[!ht]
	\caption{Monte Carlo Tree Search for TCR Generation}
	\label{alg:mcts}
	\begin{algorithmic}[1]
	\Require $\reward(., \peptide)$, $c_{puct}$, $N$, minLen, maxLen
	\State $U = \{\}$
	\For{$n=$ 1 to $N$}
	\ShortLineComment{root node of search tree $T$}
	\State $\tcr_0 = $ {``C''}; $\tau^n = \{\}$
	\For{$l=$ 1 to maxLen}                   
	\vspace{0.3em}
	\ShortShortLineComment{expansion}
	\For{each amino acid \amino}
	\State $\tcr_l' = \tcr_{l-1} + \amino$
	\If{$\tcr_l' \notin T$}
	\State $W(\tcr_l')=0$; $N(\tcr_l')=0$; $P(\tcr_l')=0$
	\State add $\tcr_l'$ as a child of $\tcr_{l-1}$ into $T$
	\EndIf
	\EndFor
	\vspace{0.3em}	
	\ShortShortLineComment{selection}
	\State select a child $\tcr_l$  of $\tcr_{l-1}$ with the maximum value $a_k = \arg \max_a \frac{W(\tcr_l)}{N(\tcr_l)} + c_{puct} * \sqrt{\frac{2N(\tcr_{l-1})}{1+N(\tcr_l)}}$
	\vspace{0.3em}
	\If{$N(\tcr_l) = $ 0 and length($\tcr_l$) $\geq$ minLen}
	\State $P(\tcr_l) = \reward(\tcr_l, \peptide)$
	\EndIf
	\vspace{0.3em}
	\State $N(\tcr_l) = N(\tcr_l) + 1$
	\State add $\tcr_l$ into $\tau^n$
	 \If{$\tcr_l$ is qualified}
 	\State \textbf{break}
 	\EndIf
 	\EndFor
 	\ShortLineComment{backpropagation}
 	\State $r^n = 0$
 	\For{each $\tcr_l$ along path $\tau^n$ from $\tcr_{\text{maxLen}}$ to $\tcr_0$}
 	\If{$P(\tcr_l) > r^n$}
 	\State $r^n = \tcr_l.P$
 	\EndIf
 	\State $W(\tcr_l) = W(\tcr_l) + r^n$
 	\EndFor
 	\ShortLineComment{output}
 	\State add $\tcr_l$ with maximum $P(\tcr_l)$ into $U$
 	\EndFor
	\vspace{0.3em}
	\State \Return $U$
	\end{algorithmic}
\end{algorithm}

\section{{Additional Results}}
\label{appendix:result}

\subsection{{Discussion on Comparison among Generation-based Methods}}
\label{appendix:result:discuss}

Generation-based methods \mcts and \bpvae on average very much underperform mutation-based methods. 
Both \mcts and \bpvae hardly generate any qualified TCRs (very low \qr: $0.01\!\pm\!0.03\%$ for \mcts and $0.04\!\pm\!0.09\%$ for \bpvae {in \mcpasP}). 
\mcts explores potentially the entire sequence space, including both valid TCRs and non-TCRs. 
Although it uses \reward to guide the search, 
with substantially more calls to calculate \reward than other methods, 
due to the fact that valid TCRs may only occupy an extreme small portion of the entire sequence space, it is extremely challenging for \mcts to 
find qualified TCRs. 
{In \vdjdbP, \mcts even cannot find any qualified TCR for all the peptides within 1,000 rollouts, and thus has zero values at $\avgistcr(\qTCRs)$ and $\avgrecog(\qTCRs)$.}
An estimation using \tcrdb over all possible length-15 sequence space results at most $\frac{1,817,721}{20^{15}} = 5.55\times 10^{-12}$\% TCRs being valid in the sequence space.
Therefore, it is possible that \mcts ends at a region in the sequence space with \qr even worse than random selection method \randomselect.

Similarly, \bpvae also has a minimum \qr ($0.04\!\pm\!0.09\%$). \bpvae encodes valid TCRs into its latent space, and uses a predictor in the latent space, 
which is 
trained to approximate \ergo, to guide the search of an optimal latent vector. This vector could correspond to a binding TCR and then is decoded 
to a TCR sequence. 
However, there is a phenomenon that is observed in other VAE-based generative approaches~\cite{GomezBombarelli2018}: while the latent vector is searched 
under a guidance to maximize desired properties, its decoded instance does not always have the properties or are not even valid. 
This phenomenon also appears in \bpvae:  the decoded TCRs are not qualified most of the times. This might be due to the propagation or
magnification of the errors from the predictor in the latent space, or the exploration in the latent space ends at a region far away from 
that of valid TCRs. However, theoretical justification behind VAE-based generative approaches is out of the scope of this paper.

\subsection{{Overview of Amino Acid Distributions of TCRs}}
\label{appendix:result:overview}

Valid TCRs, regardless of their binding affinities against any peptides, have certain patterns~\cite{Freeman2009}. For example, 
the first and last amino acids have to be `C' and `F', respectively.
Figure~\ref{fig:motif_length} presents the patterns among different TCRs of length 13, 14, 15 and 16. 
For TCRs of length 15 (most common TCR length), the left panel of Figure~\ref{fig:motif_15} shows the enrichment of different amino acids at different positions among peptides in \tcrdb.
Recall that \tcrdb has 277 million valid human TCRs, and thus the patterns in 
the left panel of Figure~\ref{fig:motif_15} are very representative of the common patterns among valid TCRs. 
The middle panel of Figure~\ref{fig:motif_15}  shows that the patterns among the TCRs that {\tcrppo}
produces for peptides in {\mcpas} are similar to those of TCRs in {\tcrdb}.
For example, `G' is one of the most frequent amino acids on positions 6-9; the first three positions and the last four positions of optimized TCRs
have patterns highly similar to those in {\tcrdb} TCRs.
Similar observations are also for the TCRs that {\tcrppo} produces for peptides in {\vdjdbP} as in the right panel of Figure~{\ref{fig:motif_15}}.
Please note that in {\tcrppo}, we do not impose any rules to reserve amino acids at the ends of sequences  
when generating TCRs with both high validity and high recognition probability against given peptides; 
clearly, {\tcrppo} successfully learns the pattern.
In Figure~\ref{fig:motif_13}, \ref{fig:motif_14} and \ref{fig:motif_16}, 
the same observations are for TCRs of length 13, 14 and 16 as in Figure~\ref{fig:motif_15} for TCRs of length 15.
For example, the first three positions and the last three positions of generated TCRs
have very similar patterns with those in {\tcrdb} (i.e., dominated by ``CAS'' and ``QYF'').
This indicates that {\tcrppo} also successfully learn the patterns for TCRs of other lengths.

\begin{figure}[!h] 
	\sidesubfloat[]{%
		\includegraphics[width=0.27\textwidth]{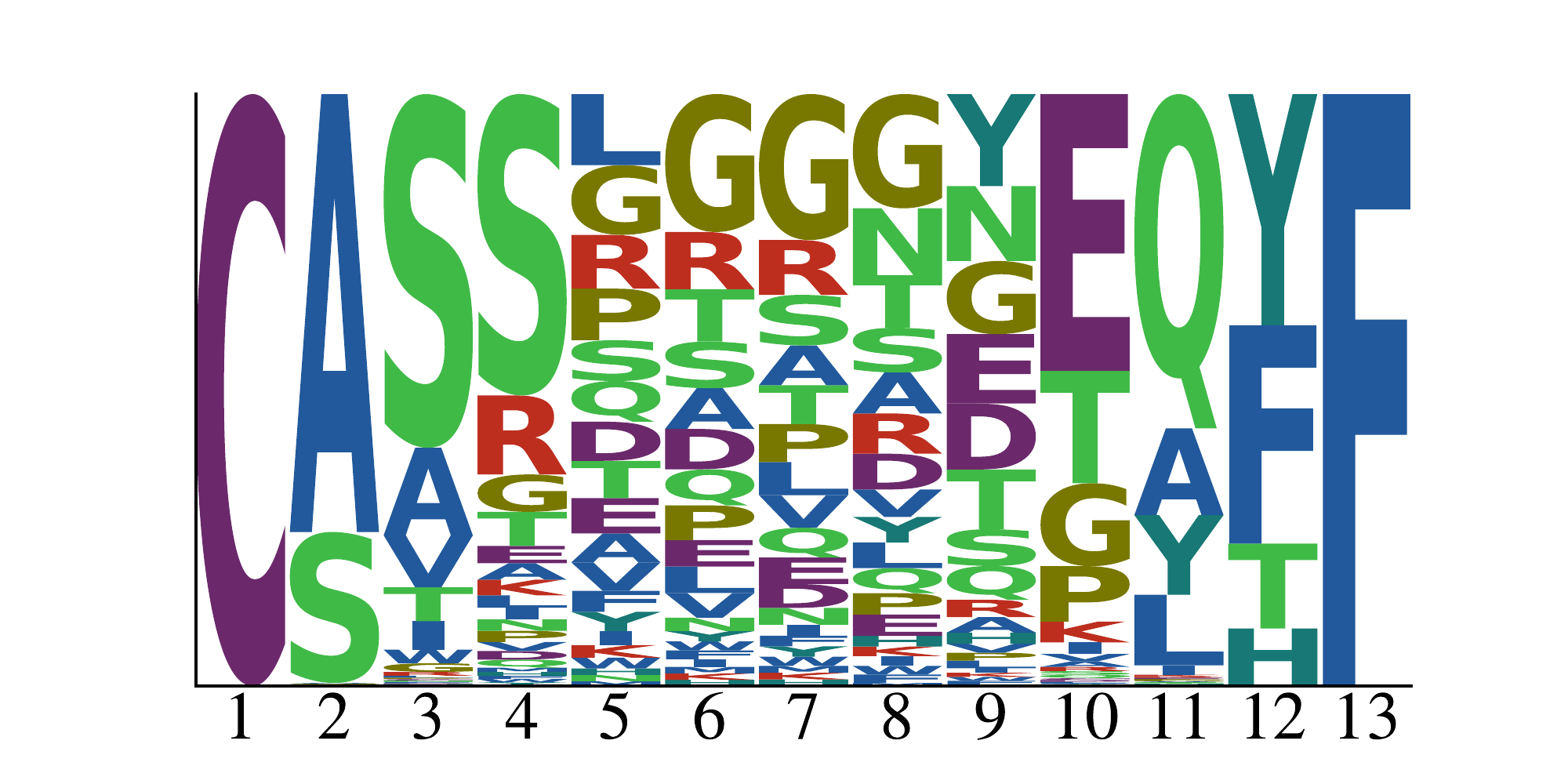}%
		\includegraphics[width=0.27\textwidth]{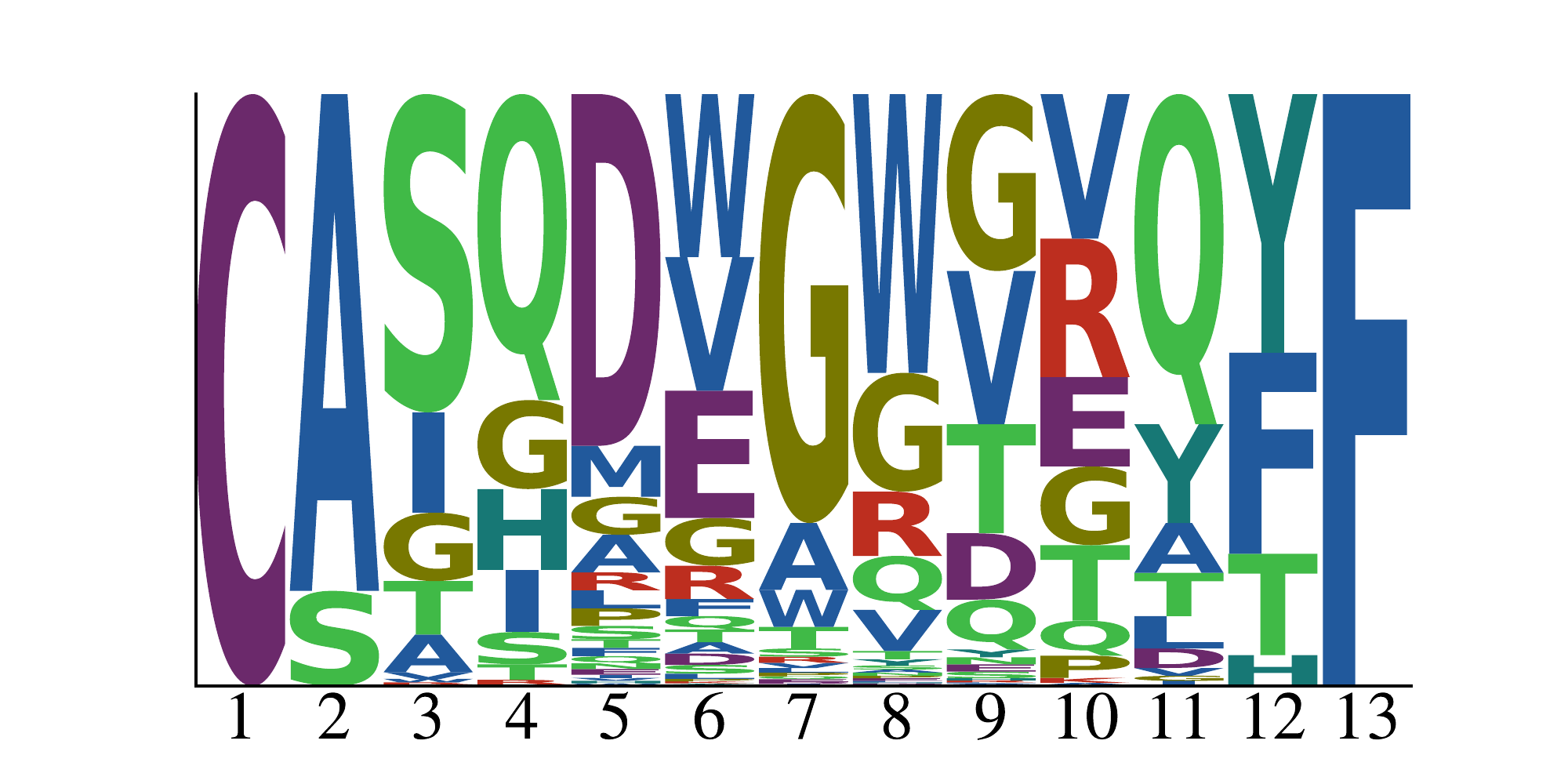}%
		\includegraphics[width=0.27\textwidth]{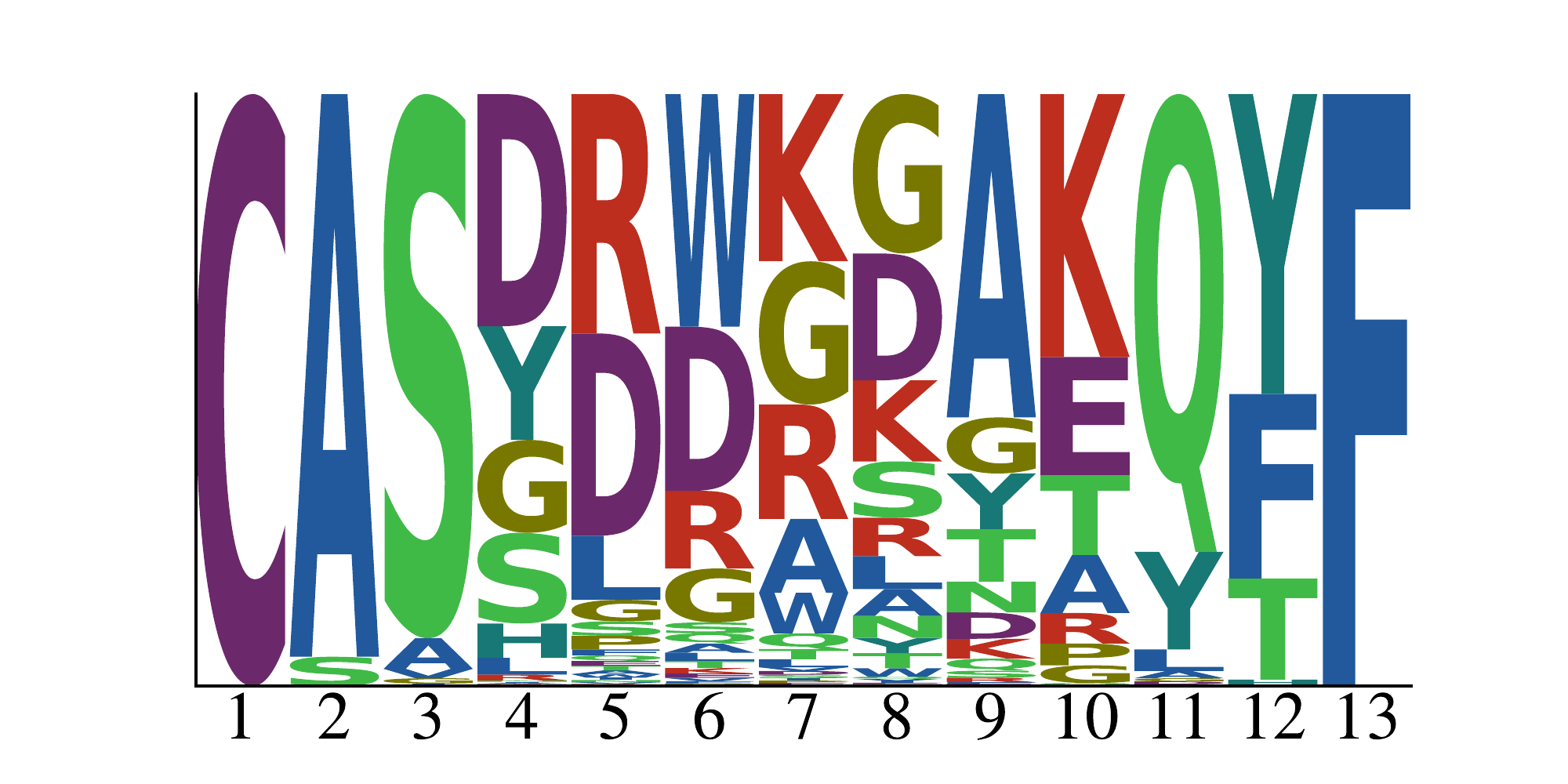}%
		\label{fig:motif_13}
	}\\
	\sidesubfloat[]{%
		\includegraphics[width=0.27\textwidth]{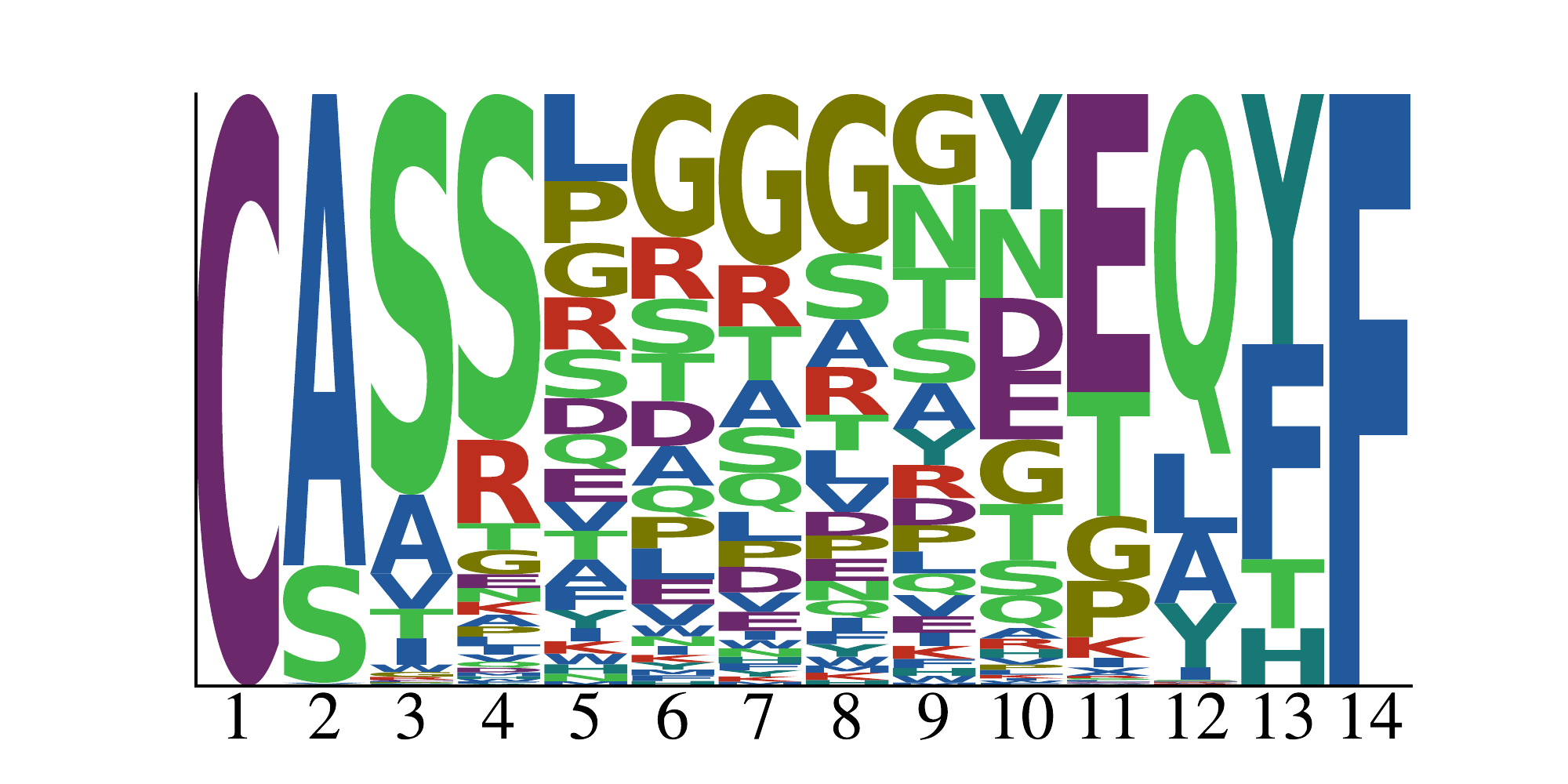}%
		\includegraphics[width=0.27\textwidth]{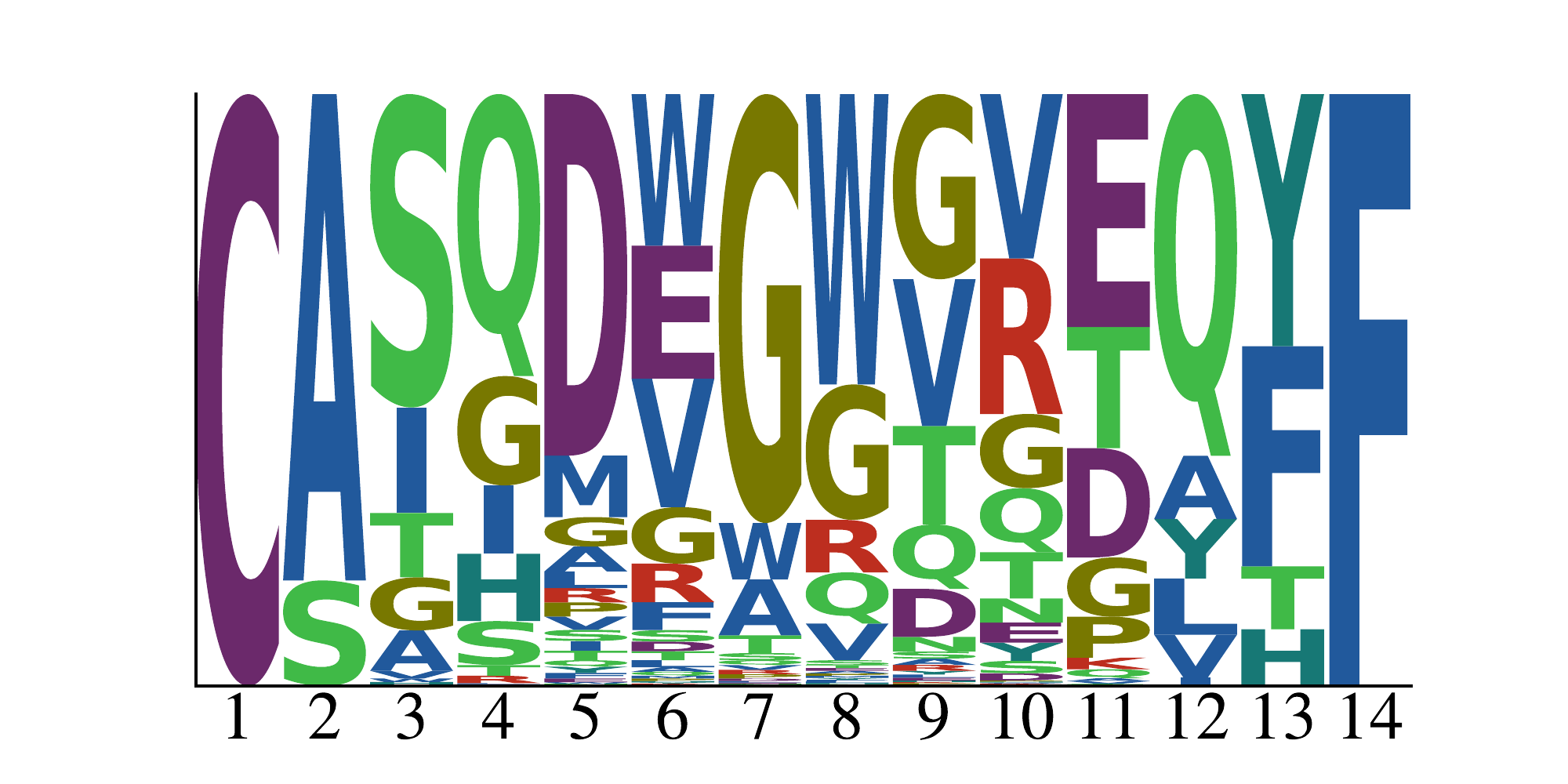}%
		\includegraphics[width=0.27\textwidth]{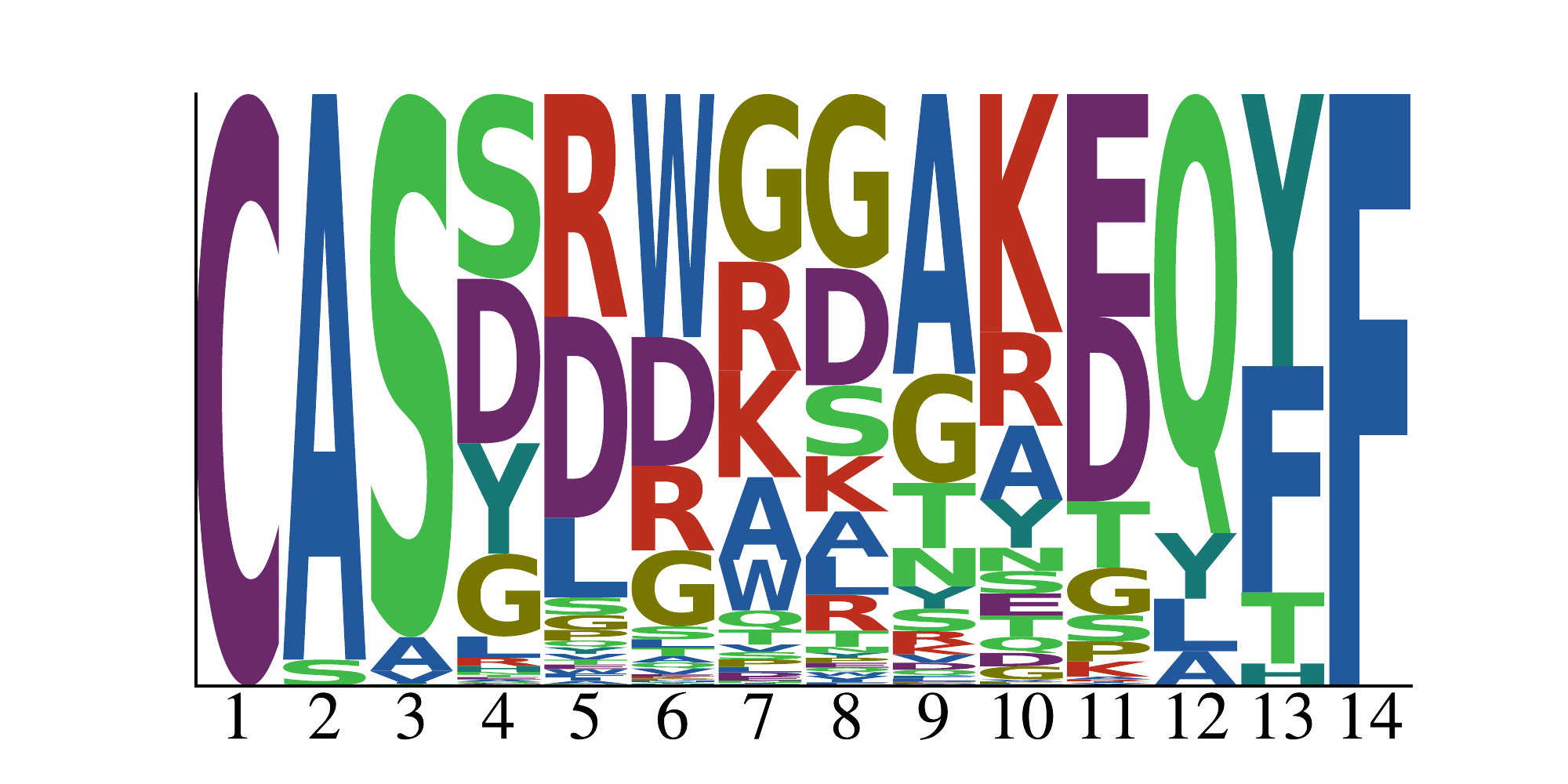}%
		\label{fig:motif_14}
	}\\
	\sidesubfloat[]{%
		\includegraphics[width=0.27\textwidth]{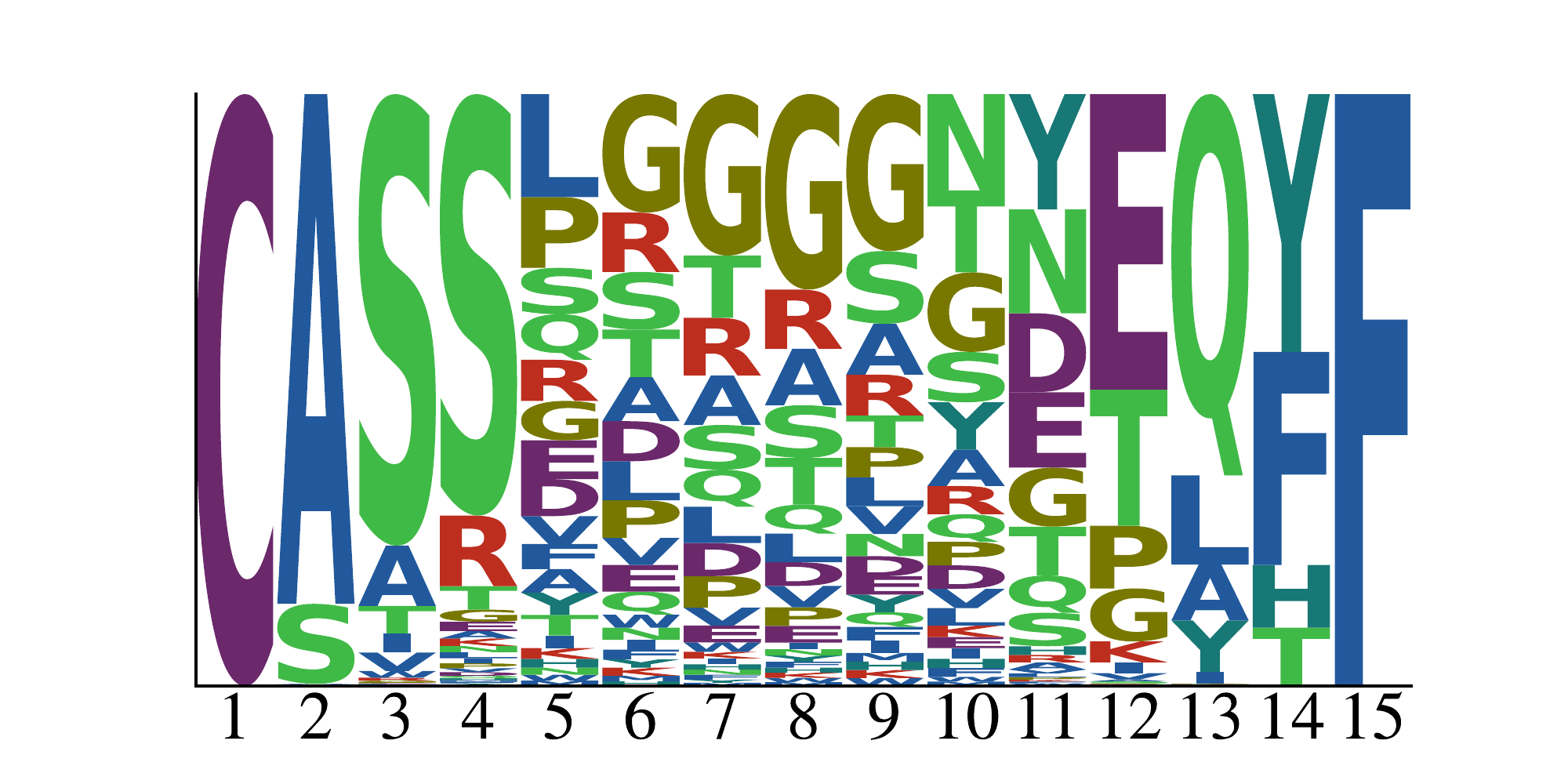}
		\includegraphics[width=0.27\textwidth]{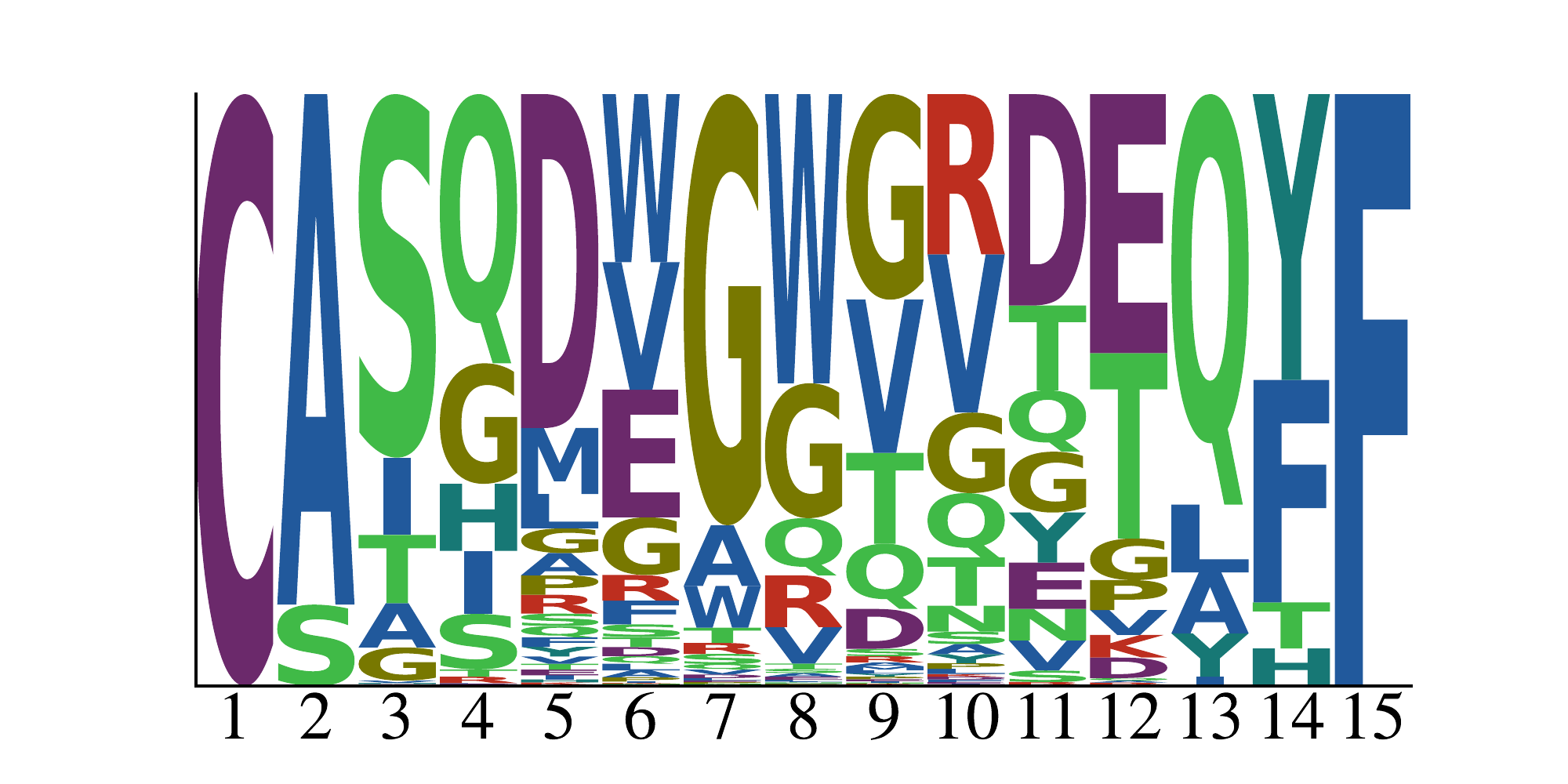}
		\includegraphics[width=0.27\textwidth]{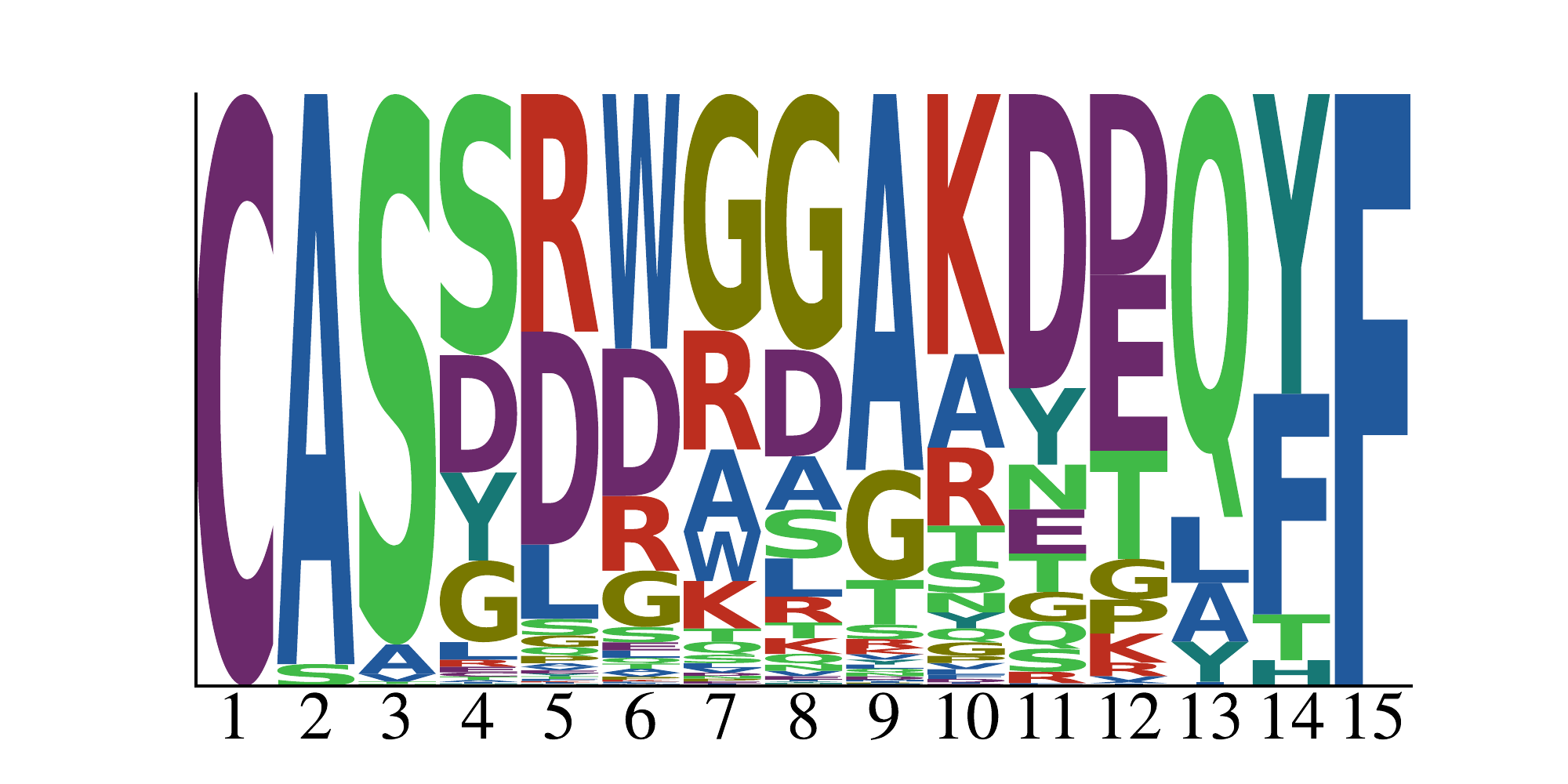}
		\label{fig:motif_15}
	}\\
	\sidesubfloat[]{%
		\includegraphics[width=0.27\textwidth]{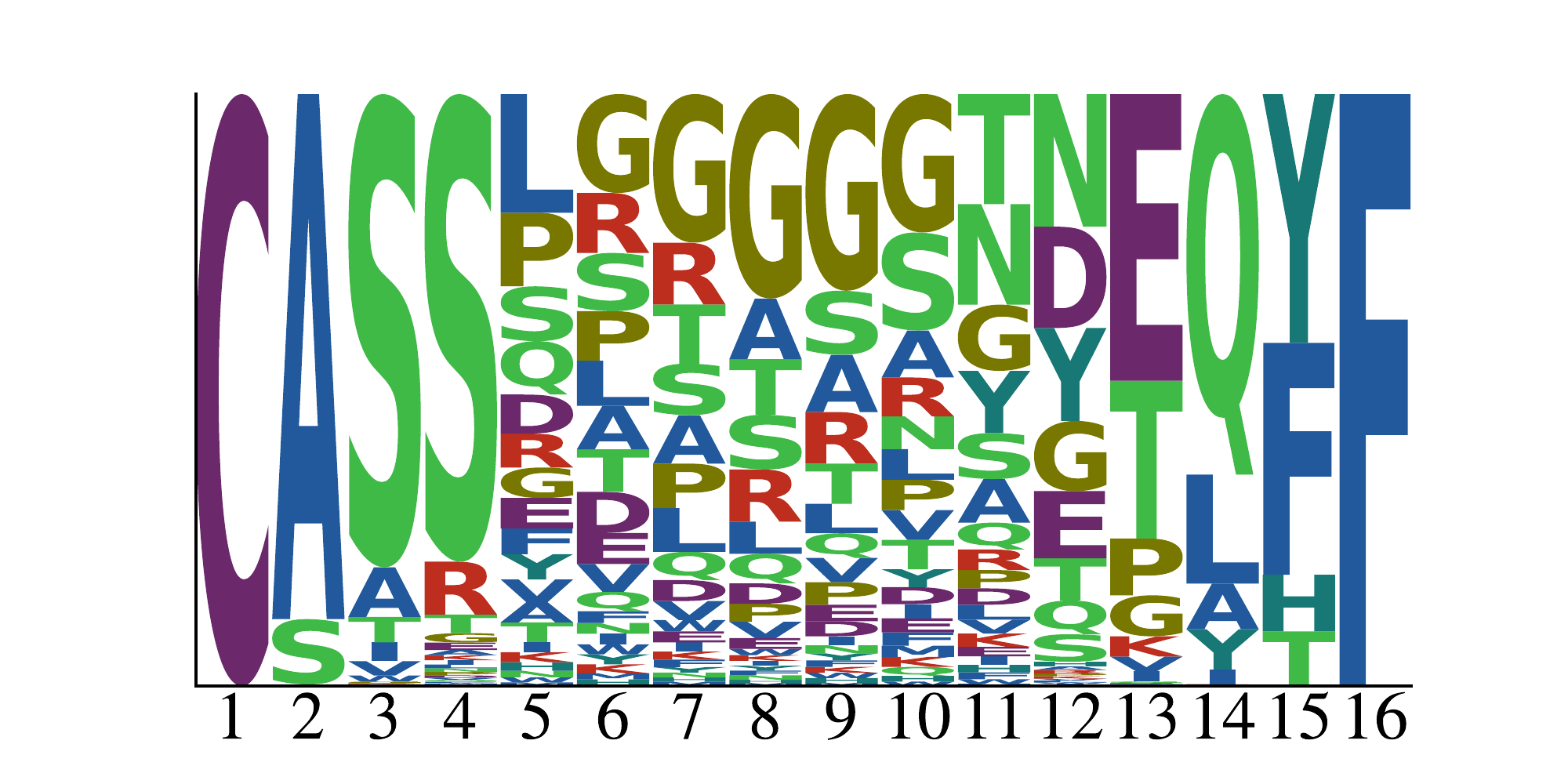}%
		\includegraphics[width=0.27\textwidth]{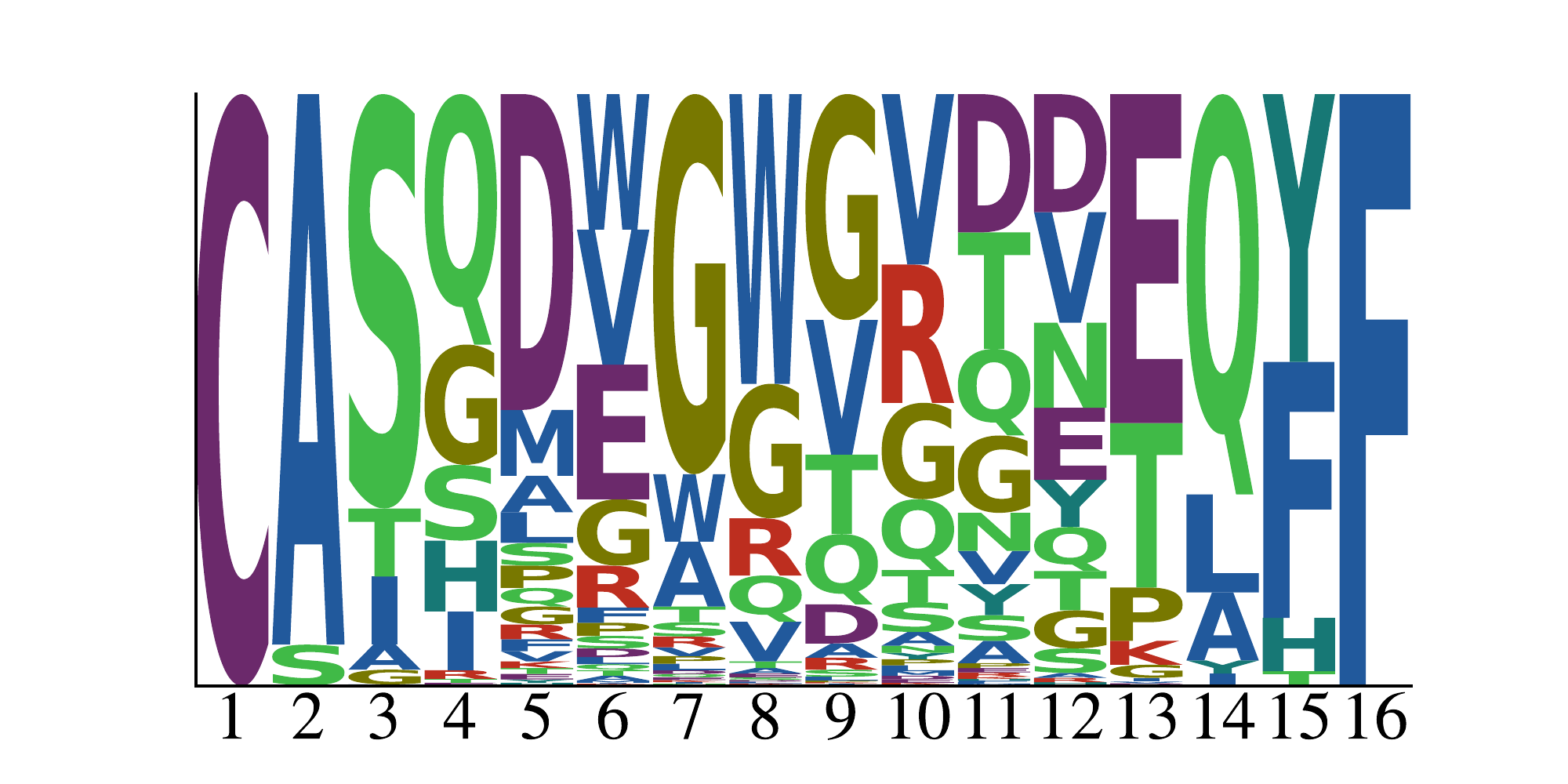}%
		\includegraphics[width=0.27\textwidth]{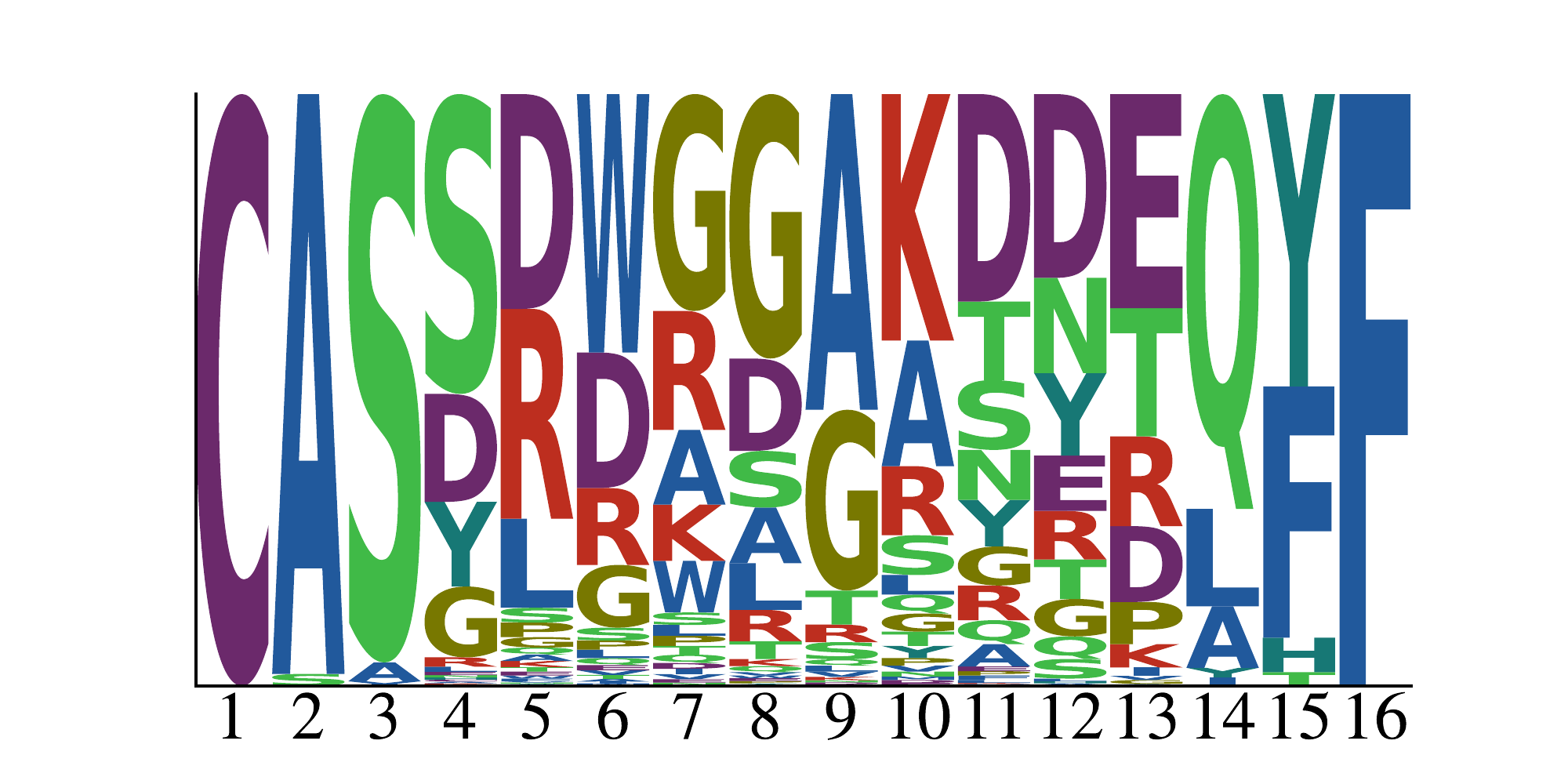}%
		\label{fig:motif_16}
	}
	\caption{{The distribution of amino acids at different positions for all the TCRs in {\tcrdb} (left panel) and the generated TCRs for {\mcpasP} (middle panel)  and {\vdjdbP}  (right panel) of length 13 (\textbf{A}), length 14 (\textbf{B}), length 15 (\textbf{C})and length 16 (\textbf{D}).}}
	\label{fig:motif_length}
\end{figure}

\subsection{{Amino Acid Distributions of Binding TCRs}}
\label{appendix:result:binding}

{Figure~\ref{fig:binding_motif} presents the patterns among binding TCRs in \mcpas dataset (left) and the generated TCRs of different lengths for peptide ``RFYKTLRAEQASQ'' in \mcpas.
Particularly, as in Figure~\ref{fig:generated_binding motif_15}, the patterns among binding TCRs in \mcpas are similar to those of generated TCRs.
For example, the most frequent amino acids at position 6 - 9 
in generated TCRs (i.e., ``R'', ``A'', ``R'', ``G'') are also one of the most frequent amino acids at corresponding positions in known binding TCRs.
Similar observations are also for generated TCRs of different lengths.
This demonstrates that \tcrppo can successfully identify the patterns of binding TCRs.
In addition to the similar patterns, we also observe the inconsistent patterns between known binding TCRs and generated TCRs.
For example, the generated TCRs are highly conserved on position 4, that is, most fourth amino acids are ``H''; this 
conservation pattern does not exist in known binding TCRs.
Please note that ``H'' also exists on position 4 among known binding TCRs as in Figure~\ref{fig:generated_binding motif_13}. 
One reason for inconsistency between generated TCRs and known binding TCRs could be that only limited percentage of binding TCRs are with predicted recognition probabilities greater than 0.9, resulting in that generated TCRs are more conserved on 
some patterns.}

{Figure~\ref{fig:action_dist} presents the distribution of action types produced by \tcrppobuf for TCRs of length 15 binding to the peptide ``RFYKTLRAEQASQ''.
As shown in Figure~\ref{fig:action_dist}, positions 5, 7, 9
in the TCRs prefer to be Histidine (``H''), Alanine (``A''), Arginine(``R'') and Glycine (``G'').
Since Alanine and Glycine are non-polar and hydrophobic, 
the hydrophobic effect tends to be strong around these positions.
On the other hand, position 4 prefer to be Histidine, 
which has a positive charge on its side chain.
Additionally, positions 5, 6, 8 in the TCRs prefer to be Arginine, 
which is hydrophilic and has a positive charge on its side chain.
This shows that the binding pocket around these positions is more electrophilic which favors the interaction with amino acids with positive charges.
}

\begin{figure}[!h] 
  	\sidesubfloat[]{%
		\includegraphics[width=0.27\textwidth]{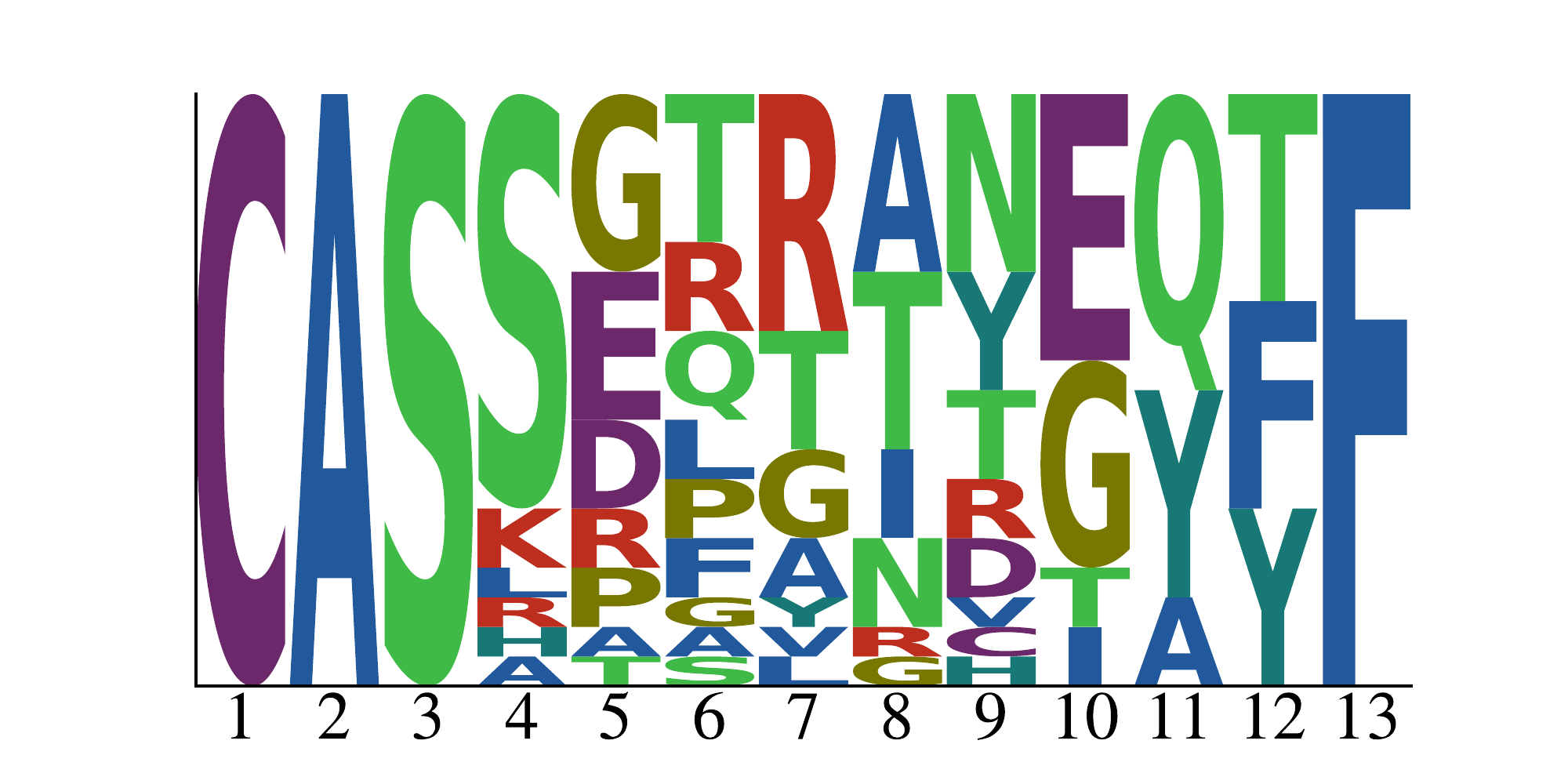}%
		\includegraphics[width=0.27\textwidth]{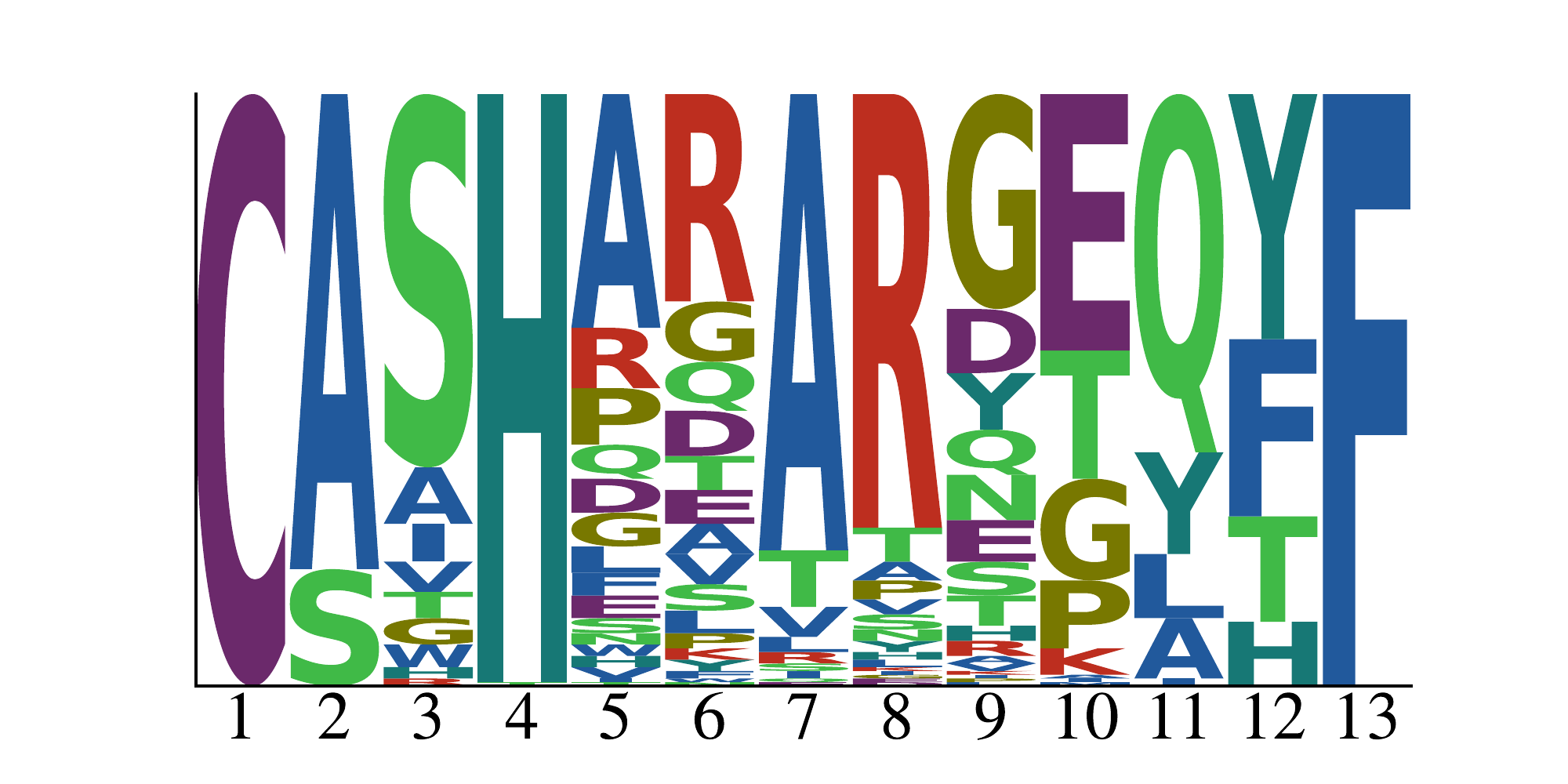}%
		\label{fig:generated_binding motif_13}
	}\\
  	\sidesubfloat[]{%
		\includegraphics[width=0.27\textwidth]{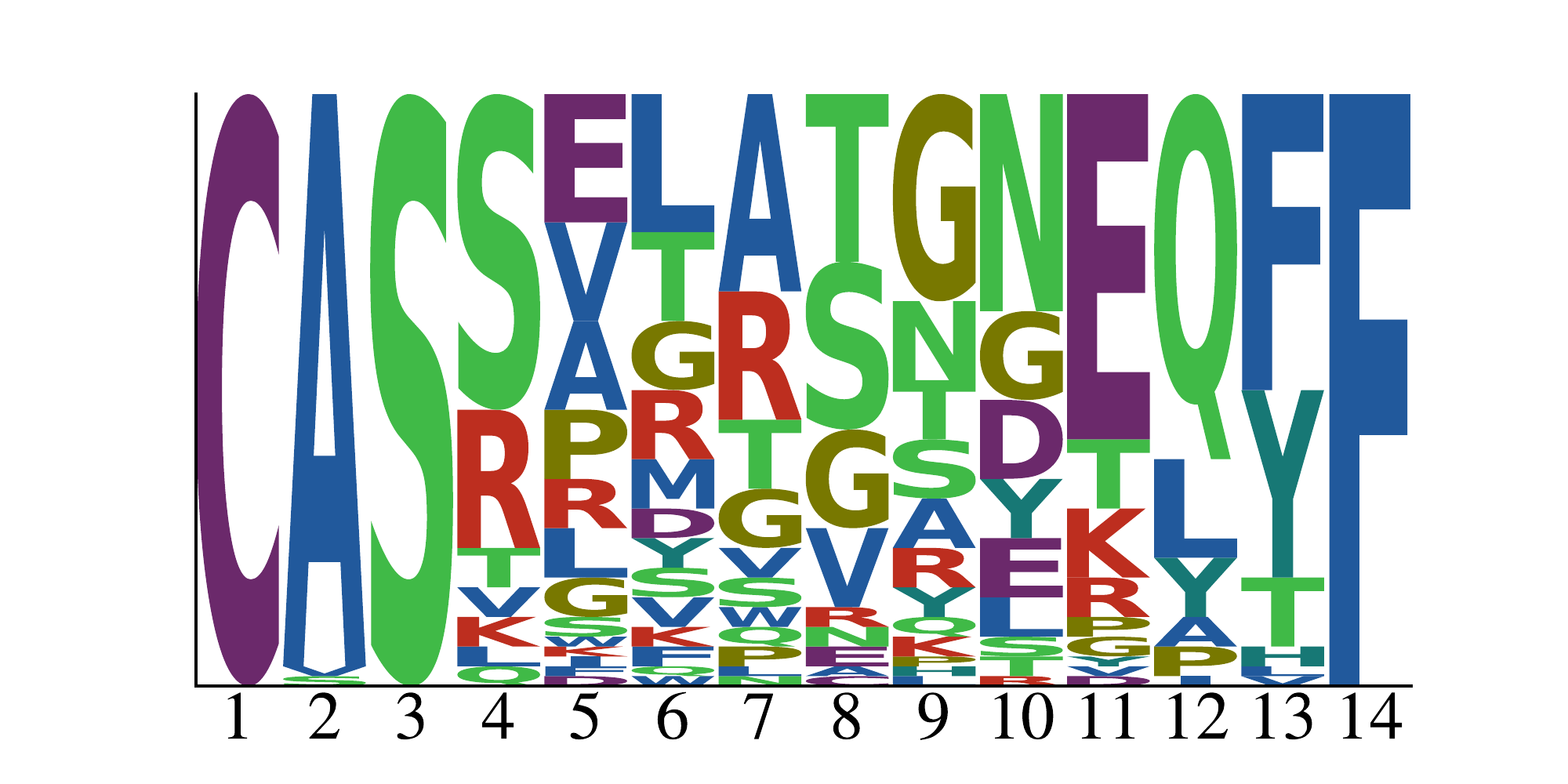}%
		\includegraphics[width=0.27\textwidth]{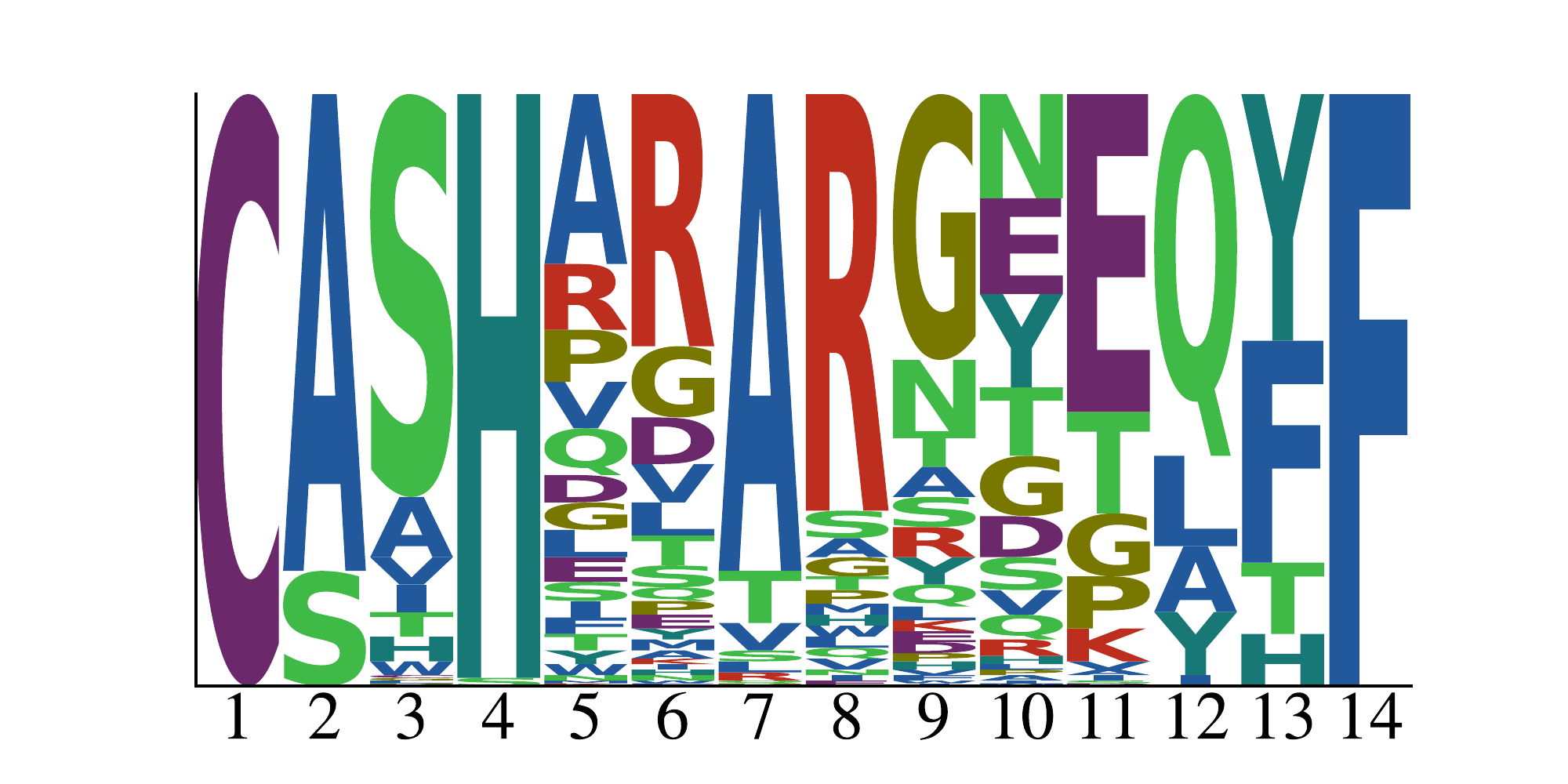}%
		\label{fig:generated_binding motif_14}
	}\\
  	\sidesubfloat[]{%
		\includegraphics[width=0.27\textwidth]{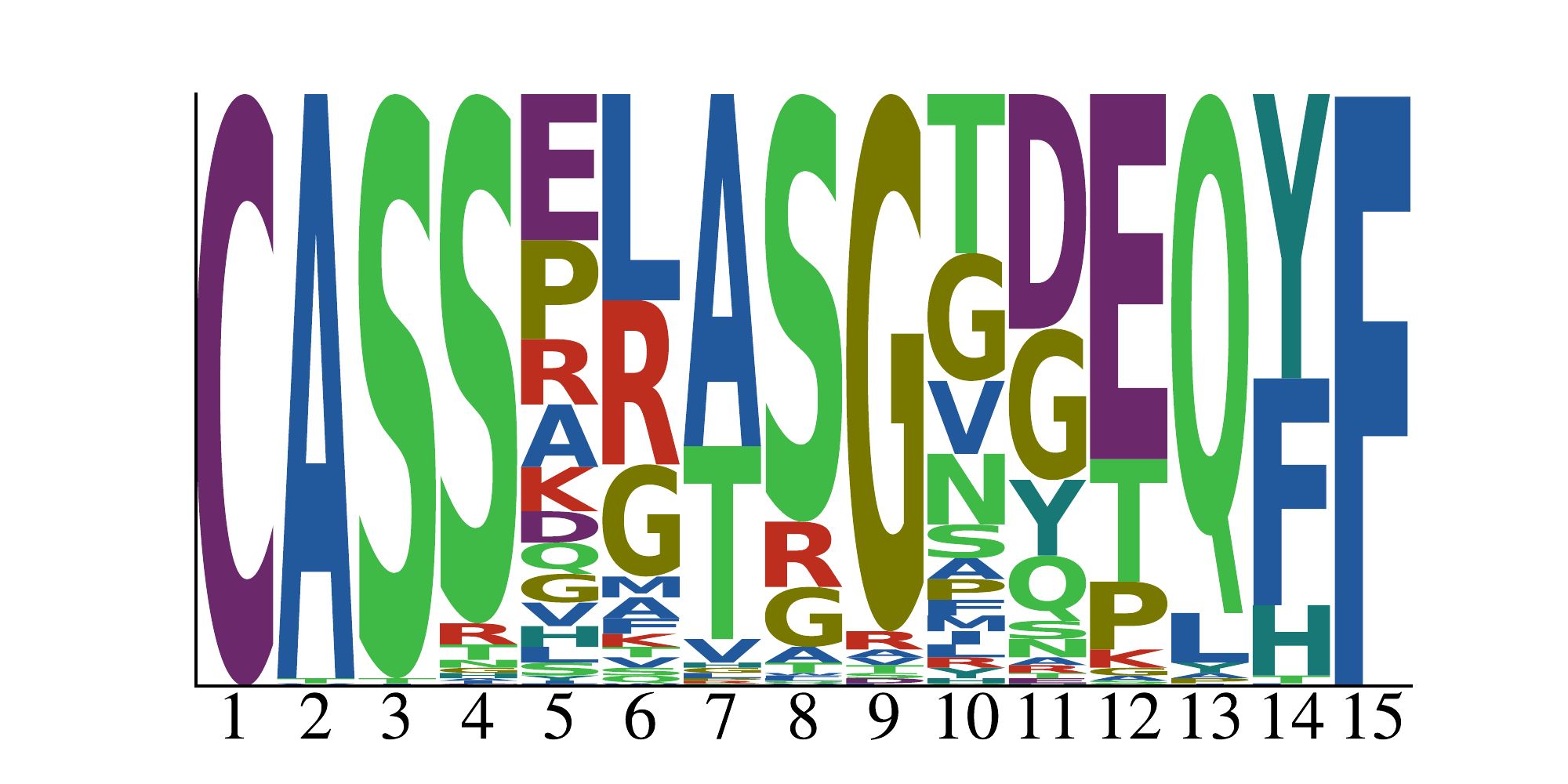}%
		\includegraphics[width=0.27\textwidth]{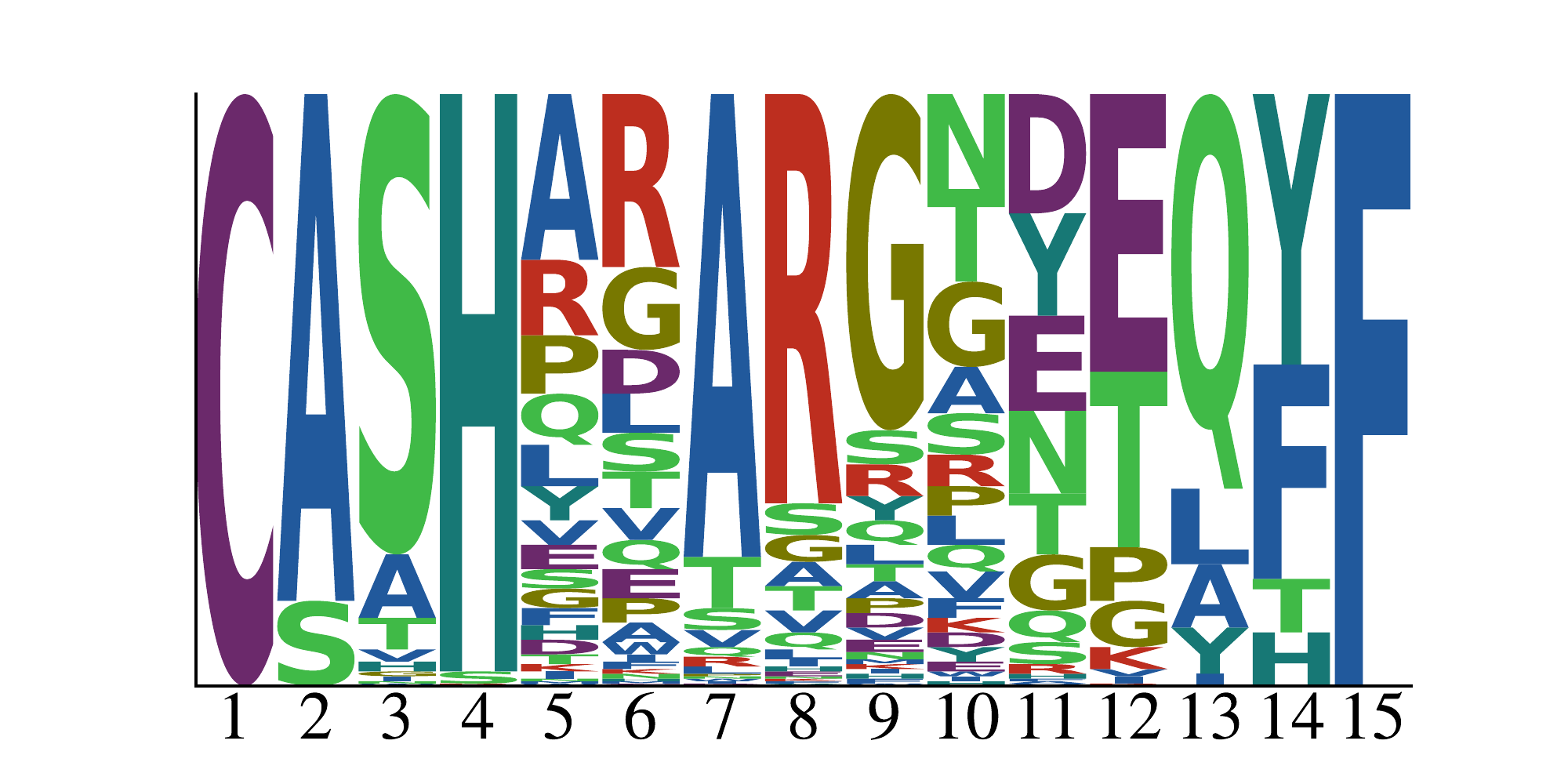}%
		\label{fig:generated_binding motif_15}
	}\\
 	\sidesubfloat[]{%
		\includegraphics[width=0.27\textwidth]{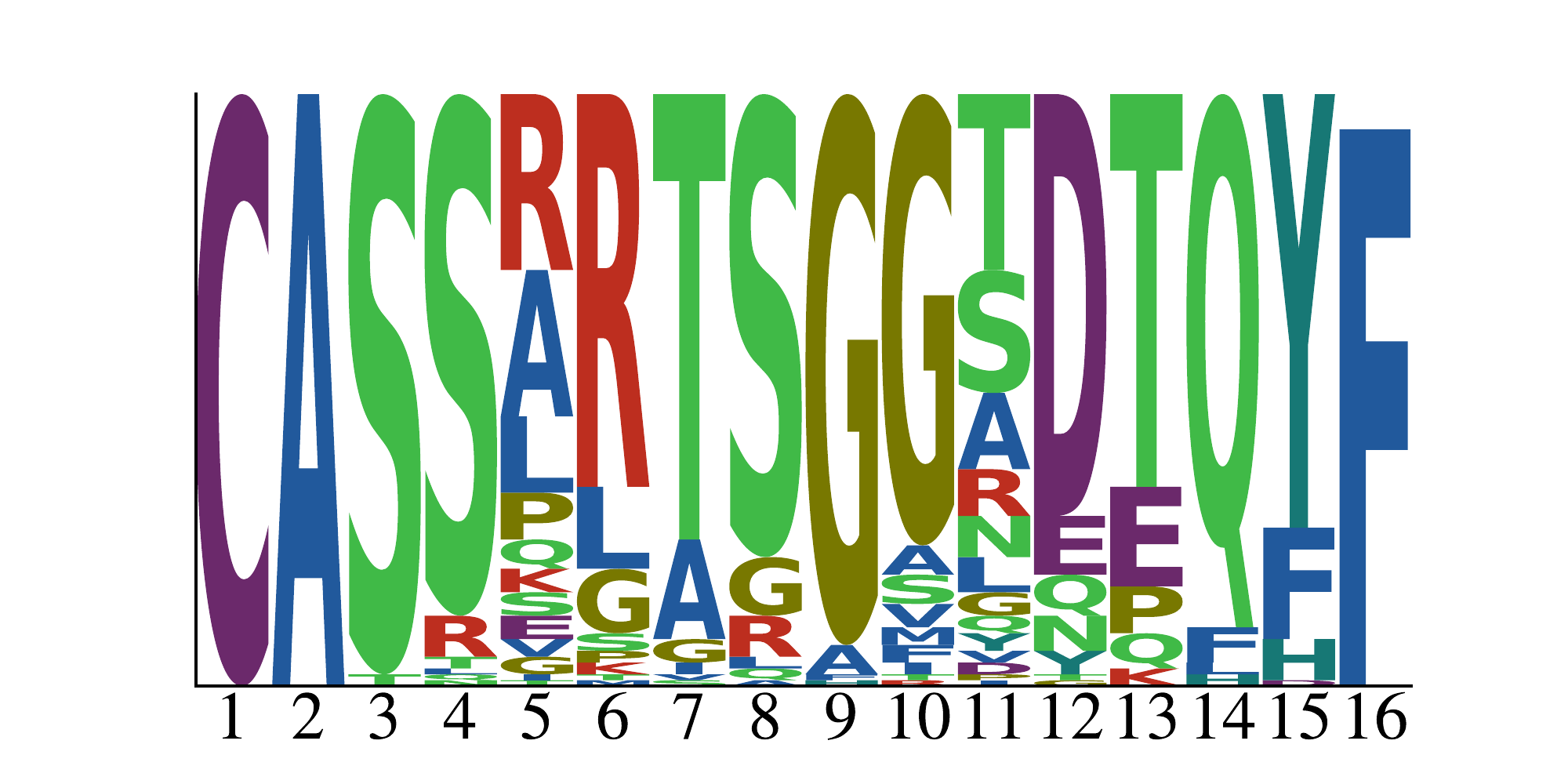}%
		\includegraphics[width=0.27\textwidth]{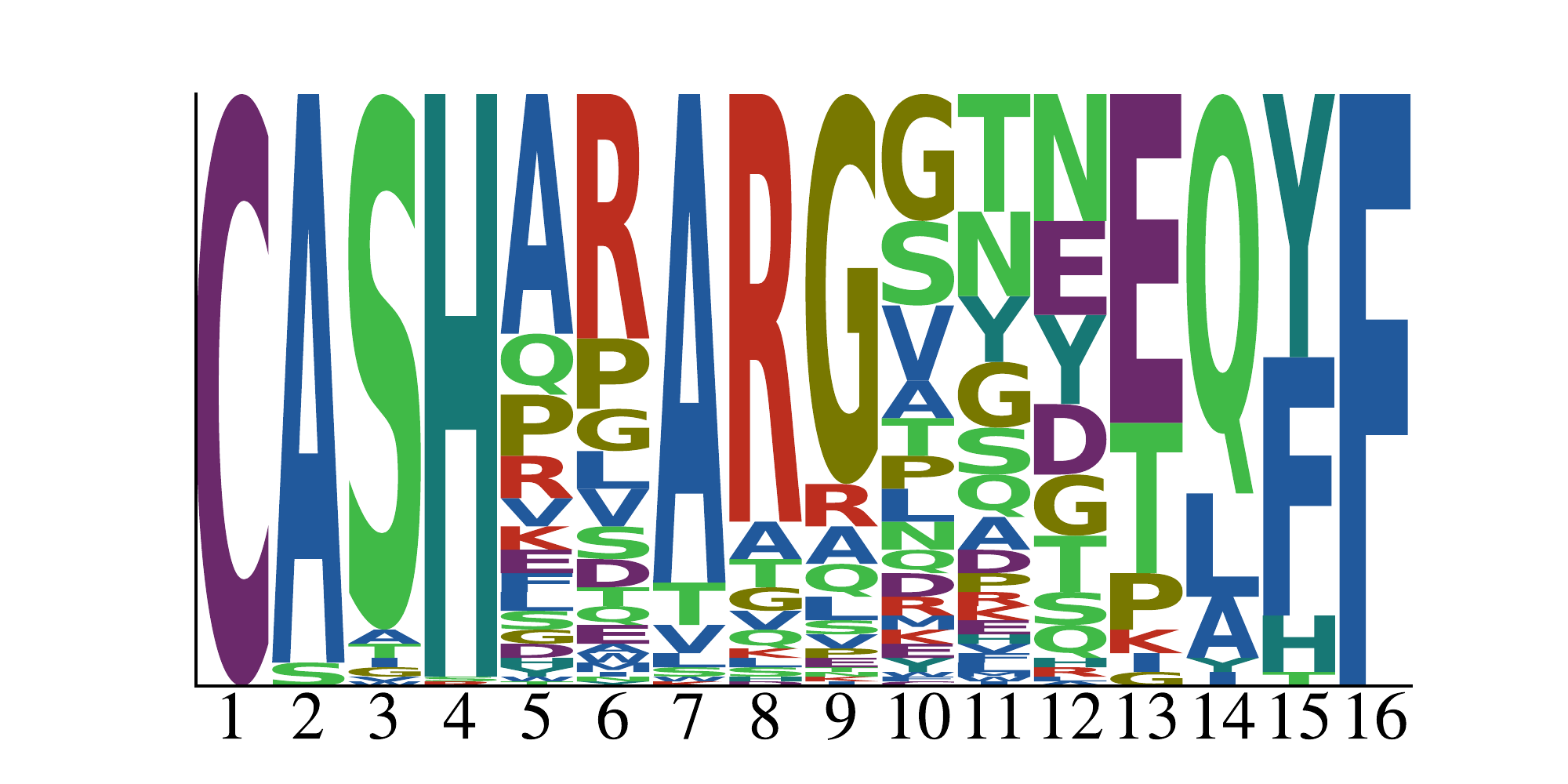}%
		\label{fig:generated_binding motif16}
	}
	\caption{{The distribution of amino acids at different positions for TCRs binding to the peptide ``RFYKTLRAEQASQ'' in \mcpas (left panel) and generated TCRs (right panel) of length 13 (\textbf{A}), length 14 (\textbf{B}), length 15 (\textbf{C}) and length 16 (\textbf{D}).}}
	\label{fig:binding_motif}
\end{figure}

\begin{figure}[!h]
\includegraphics[width=0.4\textwidth]{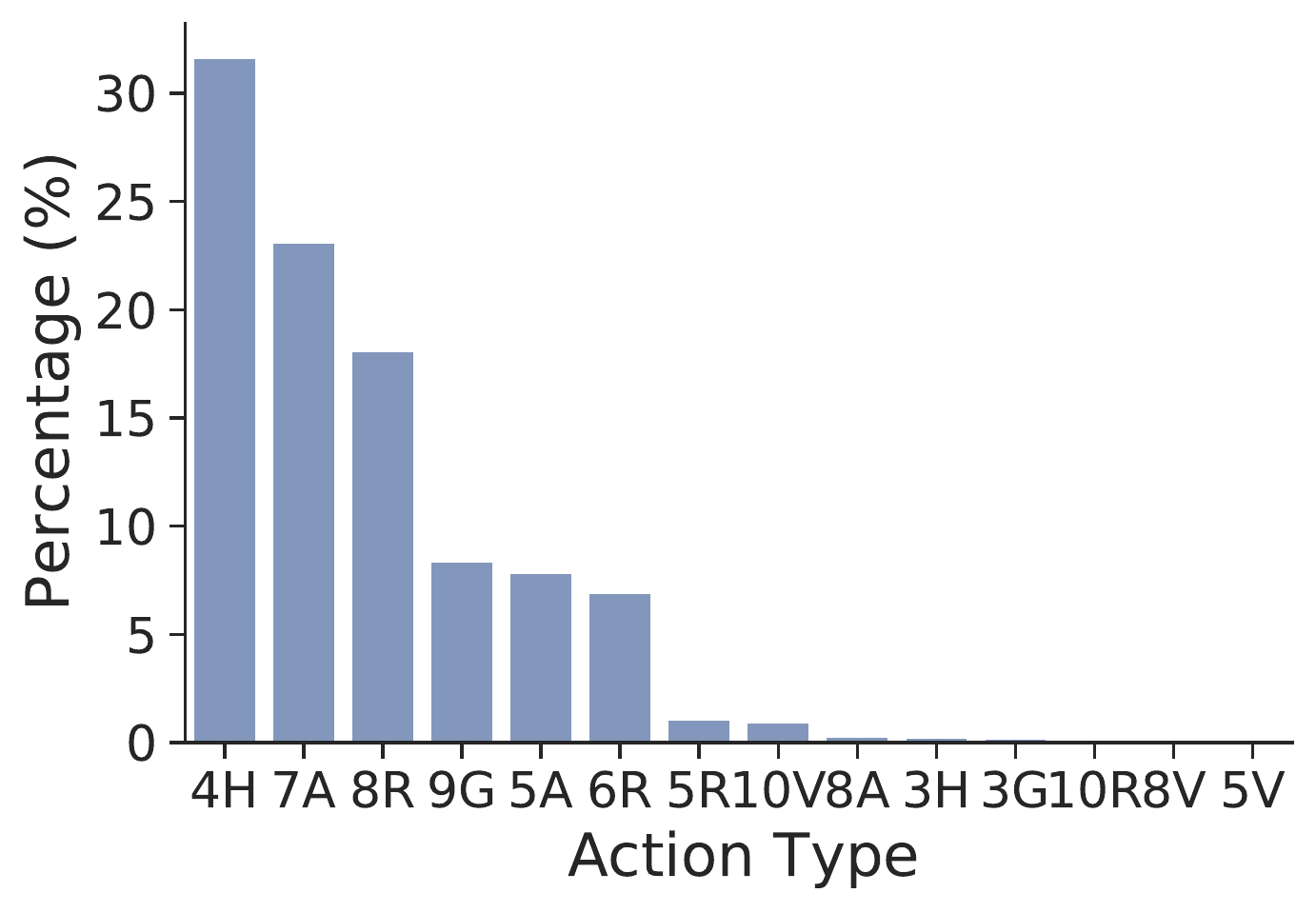}
\caption{{The distribution of action types produced by \tcrppobuf for TCRs of length 15 binding to the peptide ``RFYKTLRAEQASQ''.}}
\label{fig:action_dist}
\end{figure}

\subsection{Comparison on TCR Detection}
\label{appendix:result:reward}

We compared our \istcr scoring method (Equation~\ref{eqn:istcr}) 
and the likelihood ratio method~\cite{ren2019}, denoted as \lr, on distinguishing TCRs from non-TCRs. 
The likelihood ratio method is one of the state-of-the-art methods on out-of-distribution (OOD) detection for genomic sequences. 
To compare \istcr and \lr, 
we generated 50,000 sequences for each TCR in \Test, 
with the two ends conserved to `C' and `F' respectively as in valid TCRs, and 
all the other internal amino acids fully random. 
These generated, random sequences will be considered as non-TCRs. 
The conserved `C' and `F' ends among the non-TCRs ensure that they are not trivially separable from TCRs simply due to the ends. 
We compared \istcr and \lr scores to detect TCRs from such non-TCRs, and present their distributions in Figure~\ref{fig:reward:tcr}. 

\begin{minipage}{\linewidth}
	\centering
	\hspace{-10pt}
	\begin{minipage}{0.75\linewidth}
		\begin{figure}[H]
			\vspace{-10pt}
			\sidesubfloat[]{%
				\includegraphics[width=0.4\textwidth]{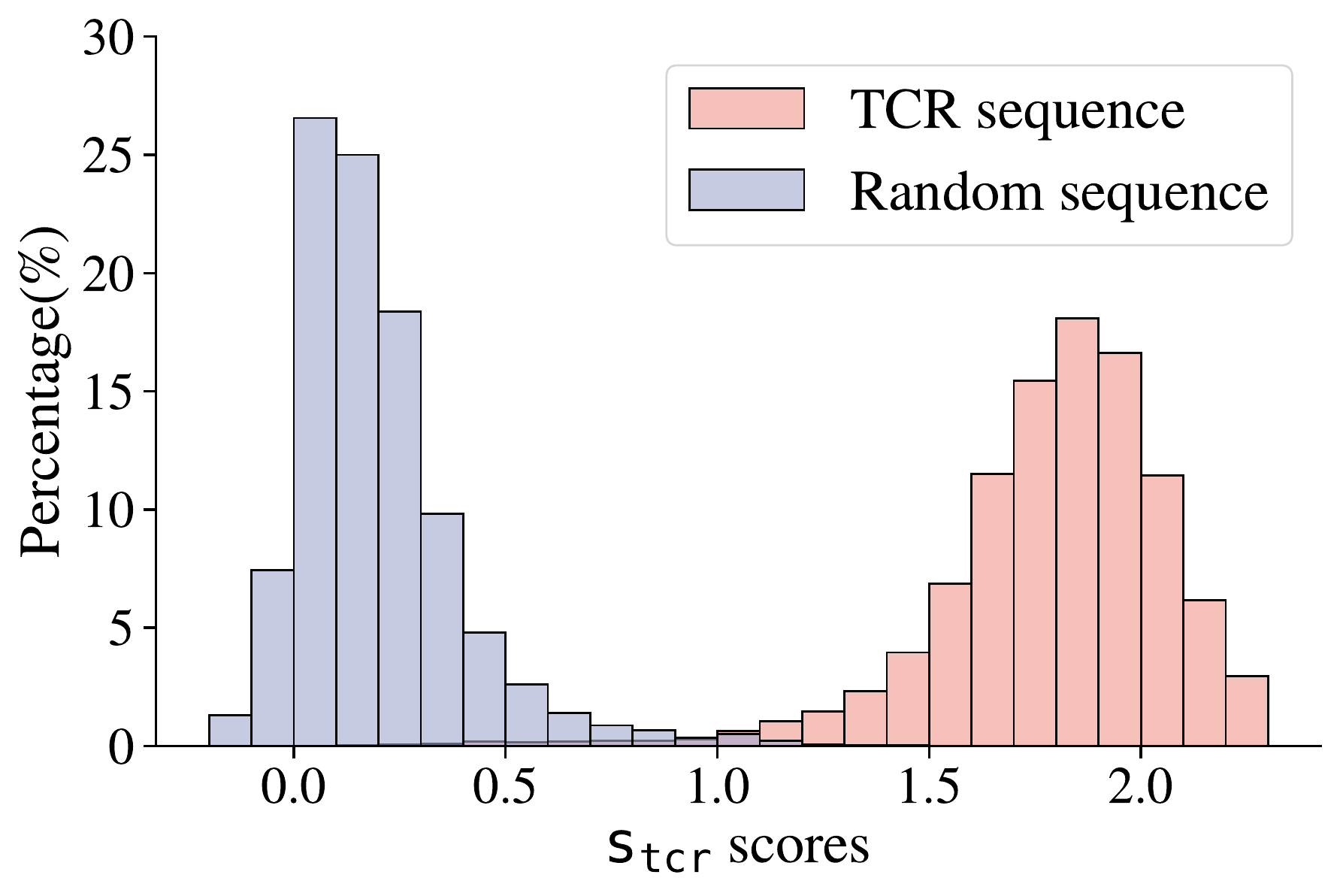}%
			}\hfil
			\sidesubfloat[]{%
				\includegraphics[width=0.4\textwidth]{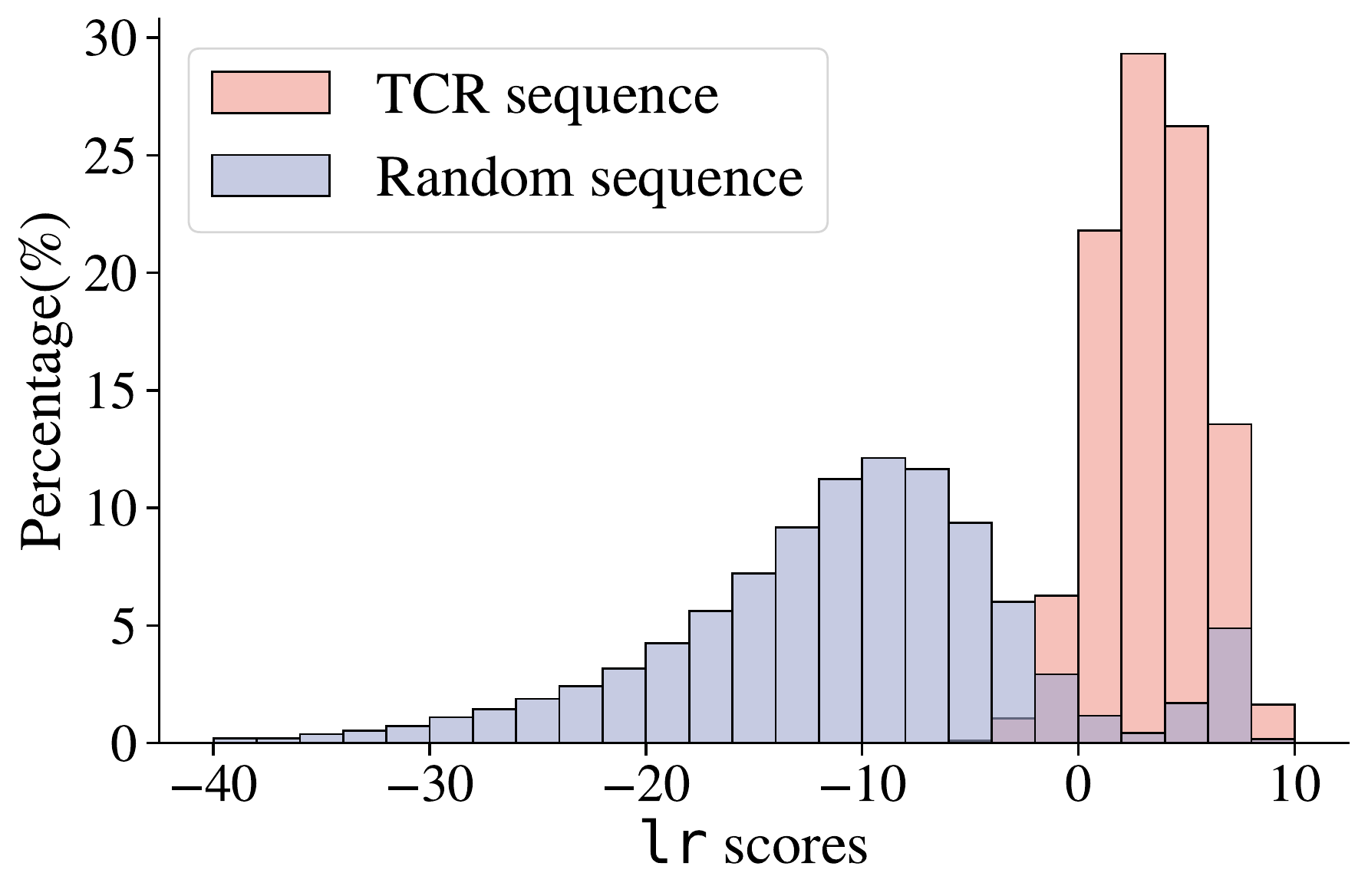}%
			}
			\caption{Distributions of \istcr scores (\textbf{A}) and \lr scores (\textbf{B}).}
			\label{fig:reward:tcr}
		\end{figure}
	\end{minipage}
	\hspace{0.01\linewidth}
	\begin{minipage}{0.75\linewidth}
		\begin{figure}[H]
		\sidesubfloat[]{%
			\includegraphics[width=0.4\textwidth]{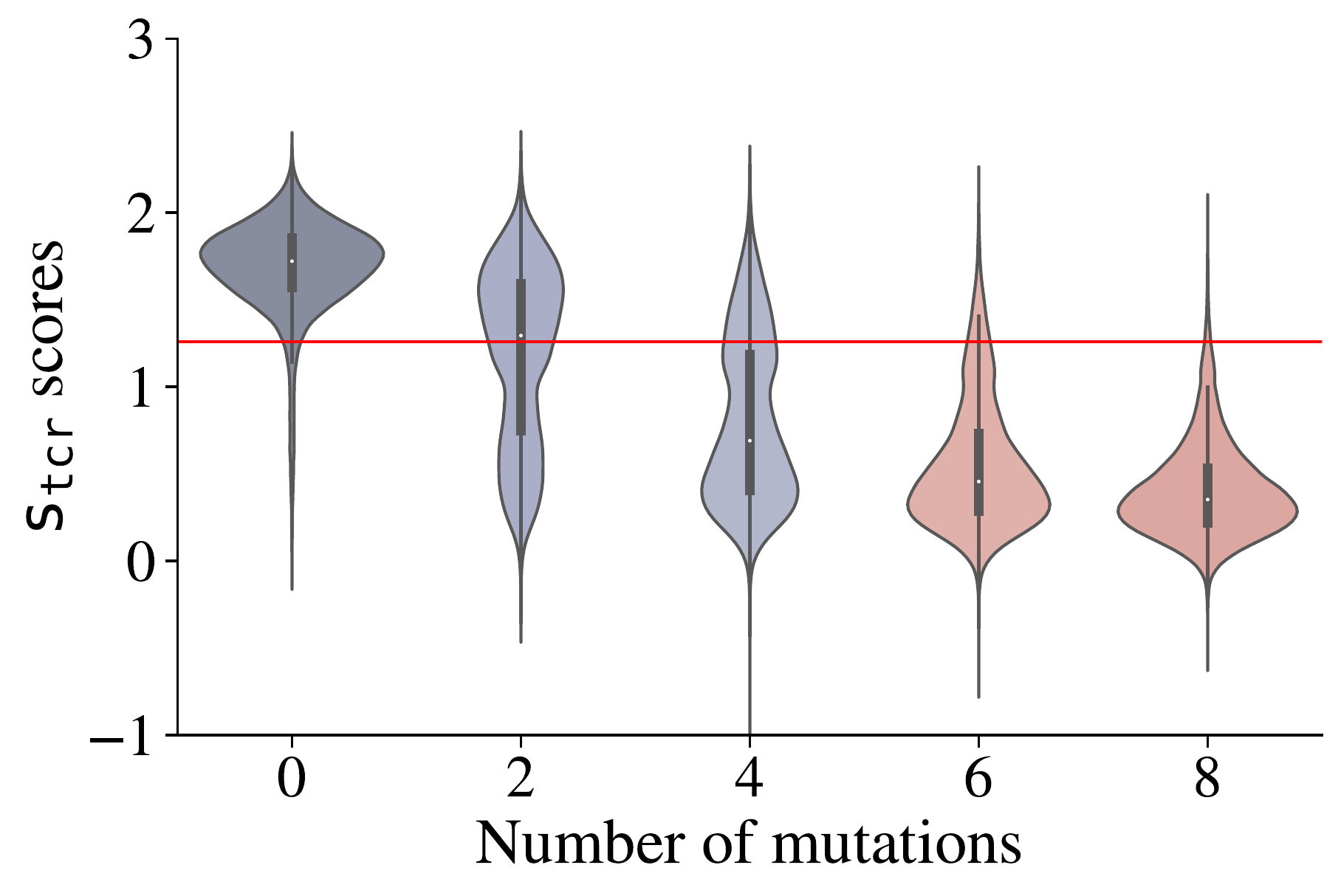}%
		}\hfil
		\sidesubfloat[]{%
			\includegraphics[width=0.4\textwidth]{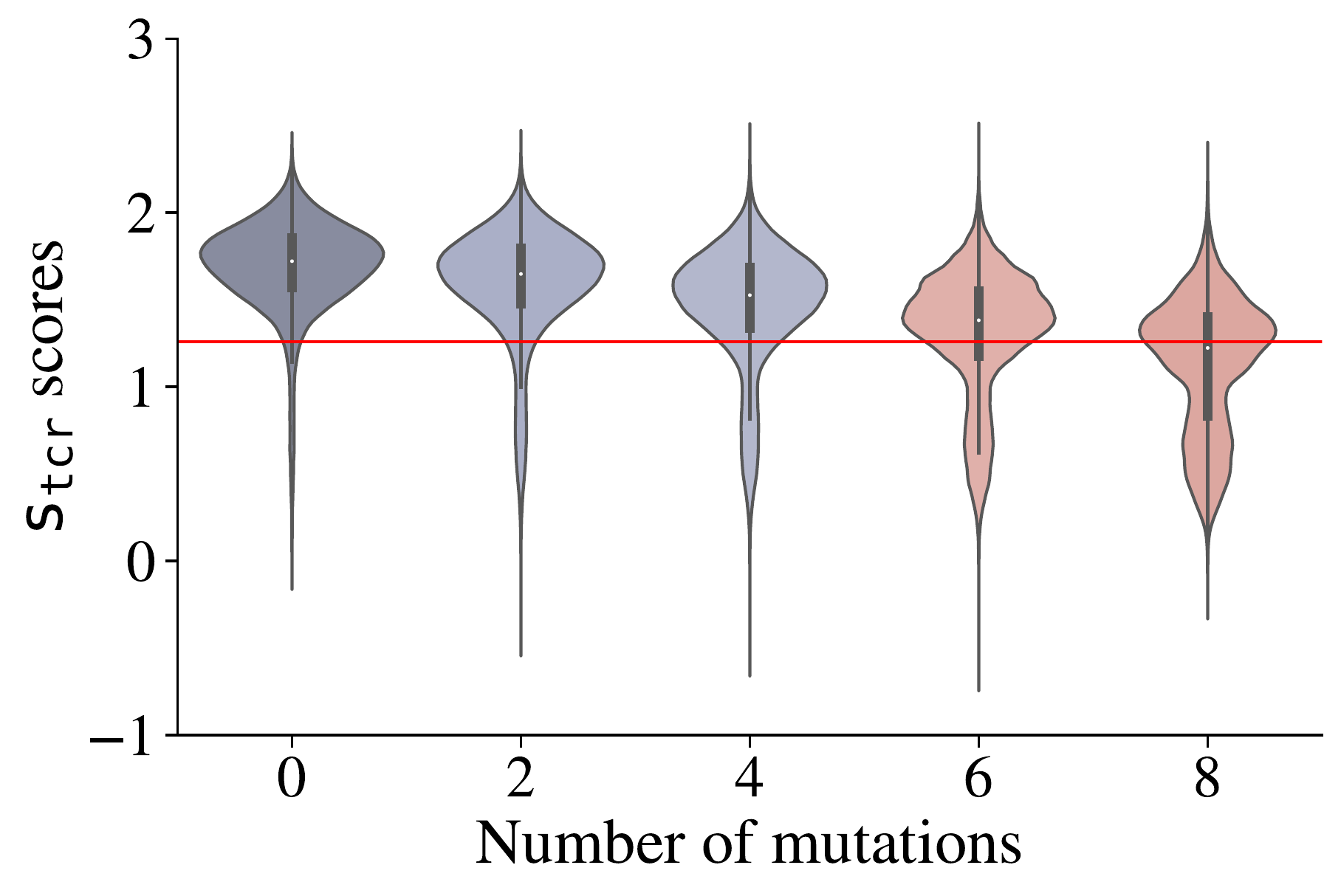}%
		}
		\caption{Distributions of \istcr scores of TCRs with random mutations (\textbf{A}) and mutations from \tcrppo (\textbf{B});
			the red lines represent the threshold value \thristcr.}
		\label{fig:reward:mut}
		\end{figure}
	\end{minipage}
\end{minipage}

Figure~\ref{fig:reward:tcr}A shows that TCRs and non-TCRs are well separated by their \istcr scores, 
while Figure~\ref{fig:reward:tcr}B shows a clear overlap between TCRs and non-TCRs using their \lr scores. 
Using the 5 percentile of \istcr scores on TCRs, corresponding to value 1.2577 and 95\% true positive rate on TCRs, as the decision boundary, 
\istcr scoring method achieves 0.10\% false positive rate. 
Using the 5 percentile of \lr scores on TCRs (corresponding to $-0.4459$), \lr has a much higher false positive rate 8.26\%. 
This demonstrates that \istcr is very effective in distinguishing TCRs. 
Please note that the random sequences are not guaranteed to be true non-TCRs. However, given the random nature of these sequences, it is highly 
likely that they are not valid TCRs. 
In \tcrppo, we used 1.2577 as the threshold to quantify if a sequence is a valid TCR (i.e., $\sigma_c = 1.2577$). 

Figure~\ref{fig:reward:mut} presents the change of \istcr scores over mutations.
Figure~\ref{fig:reward:mut}A shows that as TCRs have more random mutations, their \istcr scores decrease dramatically.
This implies that the TCR optimization via mutation is not trivial, as a random/bad mutation can significantly decrease \istcr score and easily 
lead to an unqualified solution. 
However, as Figure~\ref{fig:reward:mut}B shows, the \istcr scores do not decrease very dramatically during \tcrppo mutations, 
consistently with a good portion of mutated sequences being valid TCRs. 
This demonstrates that for this non-trivial optimization problem, \tcrppo succeeds in mutating sequences along the trajectories not far away from valid TCRs, toward final qualified sequences. 

\end{document}

%% file: define.tex
\newcommand{\etal}{\mbox{\emph{et al.}}\xspace}

\newcommand{\ergo}{\mbox{$\mathop{\mathtt{ERGO}}\limits$}\xspace}
\newcommand{\gmm}{\mbox{$\mathop{\mathtt{GMM}}\limits$}\xspace}

\newcommand{\tcrppo}{\mbox{$\mathop{\mathtt{TCRPPO}}\limits$}\xspace}
\newcommand{\tcrppobuf}{\mbox{$\mathop{\mathtt{TCRPPO}\text{+}\mathtt{b}}\limits$}\xspace}

\newcommand{\tcrdb}{\mbox{$\mathop{\mathtt{TCRdb}}\limits$}\xspace}

\newcommand{\RL}{\mbox{$\mathop{\mathtt{RL}}\limits$}\xspace}

\newcommand{\buffer}{\mbox{$\mathop{\mathtt{Buf\text{-}Opt}}\limits$}\xspace}

\newcommand{\bpvae}{\mbox{$\mathop{\mathtt{BP\text{-}VAE}}\limits$}\xspace}

\newcommand{\mcts}{\mbox{$\mathop{\mathtt{MCTS}}\limits$}\xspace}
\newcommand{\randomselect}{\mbox{$\mathop{\mathtt{RS}}\limits$}\xspace}
\newcommand{\randommutate}{\mbox{$\mathop{\mathtt{RM}}\limits$}\xspace}

\newcommand{\genetic}{\mbox{$\mathop{\mathtt{Genetic}}\limits$}\xspace}
\newcommand{\greedy}{\mbox{$\mathop{\mathtt{Greedy}}\limits$}\xspace}

\newcommand{\vect}[1]{\mathbf{#1}}

\newcommand{\tcr}{\mbox{$\mathop{c}$}\xspace}

\newcommand{\peptide}{\mbox{$\mathop{p}$}\xspace}

\newcommand{\sta}{\mbox{$\mathop{s}\limits$}\xspace}
\newcommand{\amino}{\mbox{$\mathop{o}$}\xspace}
\newcommand{\aminoemb}{\mbox{$\vect{o}$}\xspace}
\newcommand{\hidden}{\mbox{$\vect{h}$}\xspace}
\newcommand{\latent}{\mbox{$\vect{z}$}\xspace}
\newcommand{\cell}{\mbox{$\vect{c}$}\xspace}
\newcommand{\action}{\mbox{$\vect{a}$}\xspace}

\newcommand{\ncalls}{\mbox{$\mathop{\#\reward}$}\xspace}
\newcommand{\mutediff}{\mbox{$\mathop{\mathtt{edist}}$}\xspace}

\newcommand{\mcpas}{\mbox{$\mathop{\mathtt{McPAS}}\limits$}\xspace}
\newcommand{\vdjdb}{\mbox{$\mathop{\mathtt{VDJDB}}\limits$}\xspace}

\newcommand{\statespace}{\mbox{$\mathcal{S}$}\xspace}
\newcommand{\actionspace}{\mbox{$\mathcal{A}$}\xspace}
\newcommand{\reward}{\mbox{$\mathcal{R}$}\xspace}
\newcommand{\transit}{\mbox{$\mathcal{P}$}\xspace}

\newcommand{\Validation}{\mbox{$\mathop{\mathcal{S}_v}\limits$}\xspace}
\newcommand{\Train}{\mbox{$\mathop{\mathcal{S}_{trn}}\limits$}\xspace}
\newcommand{\Test}{\mbox{$\mathop{\mathcal{S}_{tst}}\limits$}\xspace}
\newcommand{\mcpasP}{\mbox{$\mathop{\mathcal{P}_{\scriptsize{\mcpas}}}\limits$}\xspace}
\newcommand{\vdjdbP}{\mbox{$\mathop{\mathcal{P}_{\scriptsize{\vdjdb}}}\limits$}\xspace}

\newcommand{\istcrsigma}{\mbox{$\mathop{\sigma_c}\limits$}\xspace}
\newcommand{\prsigma}{\mbox{$\mathop{\sigma_r}\limits$}\xspace}

\newcommand{\qr}{\mbox{$\mathop{\mathtt{q\%}}\limits$}\xspace}
\newcommand{\vr}{\mbox{$\mathop{\mathtt{v\%}}\limits$}\xspace}
\newcommand{\lr}{\mbox{$\mathop{\mathtt{lr}}\limits$}\xspace}
\newcommand{\istcr}{\mbox{$\mathop{s_v}\limits$}\xspace}
\newcommand{\thristcr}{\mbox{$\mathop{\sigma_v}\limits$}\xspace}
\newcommand{\recog}{\mbox{$\mathop{s_r}$}\xspace}
\newcommand{\reconst}{\mbox{$\mathop{r_r}$}\xspace}
\newcommand{\dense}{\mbox{$\mathop{r_d}$}\xspace}

\newcommand{\vTCRs}{\mbox{$\mathop{\mathcal{C}_v}\limits$}\xspace}

\newcommand{\qTCRs}{\mbox{$\mathop{\mathcal{C}_q}\limits$}\xspace}

\newcommand{\avgistcr}{\mbox{$\mathop{\overline{\istcr}}\limits$}\xspace}
\newcommand{\avgrecog}{\mbox{$\mathop{\overline{\recog}}\limits$}\xspace}

\newcommand{\autoenc}{\mbox{$\mathop{\mathtt{TCR\text{-}AE}}$}\xspace}

\newcommand{\ziqi}[1]{}

%% file: tables/result_full_new.tex
\begin{sidewaystable}
	\caption{Overall Performance Comparison}
	\label{tbl:perform:all}
	\centering
	\begin{footnotesize}
		\begin{threeparttable}
			\begin{tabular}{
					@{\hspace{5pt}}l@{\hspace{5pt}} 
					@{\hspace{2pt}}l@{\hspace{2pt}} 
					@{\hspace{2pt}}r@{\hspace{0pt}} 
					@{\hspace{0pt}}c@{\hspace{0pt}}
					@{\hspace{0pt}}l@{\hspace{2pt}}
					@{\hspace{2pt}}r@{\hspace{0pt}} 
					@{\hspace{0pt}}c@{\hspace{0pt}}
					@{\hspace{0pt}}r@{\hspace{2pt}}
					@{\hspace{2pt}}r@{\hspace{0pt}} 
					@{\hspace{0pt}}c@{\hspace{0pt}}
					@{\hspace{0pt}}l@{\hspace{2pt}}
					@{\hspace{2pt}}r@{\hspace{0pt}} 
					@{\hspace{0pt}}c@{\hspace{0pt}}
					@{\hspace{0pt}}l@{\hspace{2pt}}
					@{\hspace{2pt}}r@{\hspace{0pt}} 
					@{\hspace{0pt}}c@{\hspace{0pt}}
					@{\hspace{0pt}}l@{\hspace{2pt}}
					@{\hspace{2pt}}r@{\hspace{0pt}} 
					@{\hspace{0pt}}c@{\hspace{0pt}}
					@{\hspace{0pt}}l@{\hspace{2pt}}
					@{\hspace{2pt}}r@{\hspace{0pt}} 
					@{\hspace{0pt}}c@{\hspace{0pt}}
					@{\hspace{0pt}}l@{\hspace{2pt}}
					@{\hspace{0pt}}r@{\hspace{2pt}} 
				}
				\toprule
                {Dataset}
                & {Method} 
				& \multicolumn{3}{c}{{\qr}} 
				&  \multicolumn{3}{c}{{\mutediff}} 
				& \multicolumn{3}{c}{\avgistcr(\qTCRs)}   & \multicolumn{3}{c}{\avgistcr(\vTCRs)}
				& \multicolumn{3}{c}{\avgrecog(\qTCRs)} & \multicolumn{3}{c}{\avgrecog(\vTCRs)} 
				& \multicolumn{3}{c}{\vr}
				& \ncalls \\
				%
				\midrule        
				\multirow{9}{*}{\mcpasP}
				& \randomselect   &  0.05  &  $\pm$  &  0.12  
									&     &  -  &    
									&  1.46  &  $\pm$  &  0.16  
									&  \textbf{1.73}  &  $\pm$  &  \textbf{0.17}  
									&  0.95  &  $\pm$  &  0.01  
									&  0.05  &  $\pm$  &  0.00  
									&  95.22  &  $\pm$  &  0.45  
				& \textbf{1}\\
				\cmidrule{2-24}						
				& \mcts   &  0.01  &  $\pm$  &  0.03  
						   &    &  -  &    
							 &  1.38  &  $\pm$  &  0.00  
							&  1.34  &  $\pm$  &  0.03  
							&  0.93  &  $\pm$  &  0.00  
							&  0.09  &  $\pm$  &  0.09  
							&  0.59  &  $\pm$  &  0.28   
				& 200\\
			    & \bpvae   &  0.04  &  $\pm$  &  0.09  
						   &     &  -  &    
							 &  1.31  &  $\pm$  &  0.02  
							&  1.34  &  $\pm$  &  0.02  
							&  0.95  &  $\pm$  &  0.02  
							&  0.09  &  $\pm$  &  0.04 
							&  6.86  &  $\pm$  &  2.52    
				& 7\\
				%
				\cmidrule{2-24}
				& \randommutate (n=5)   &  1.27  &  $\pm$  &  2.09  
								&  \textbf{2.88}  &   $\pm$   & \textbf{1.50}
								 &  1.49  &  $\pm$  &  0.07  
								&  1.54  &  $\pm$  & 0.02  
								&  0.93  &  $\pm$  &  0.02  
								&  0.18  &  $\pm$  &  0.09  
								&  99.36  &  $\pm$  &  0.21   
				& 39\\
				& \randommutate (n=10)  &  2.15  &  $\pm$  &  3.86  
								&  3.04  &   $\pm$   & 1.48 
								 &  1.50  &  $\pm$  &  0.11  
								&  1.52  &  $\pm$  &  0.02  
								&  0.93  &  $\pm$  &  0.02  
								&  0.26  &  $\pm$  &  0.11  
								&  99.84  &  $\pm$  &  0.12   
				& 77\\
				& \greedy   &  20.58  &  $\pm$  &  17.85  
								&  5.10   &  $\pm$  & 1.09
								 &  1.46  &  $\pm$  &  0.03  
								&  1.44  &  $\pm$  &  0.02 
								&  0.93  &  $\pm$  &  0.01  
								&  0.62  &  $\pm$  &  0.12  
								&  99.98  &  $\pm$  &  0.04   
				& 74\\
				& \genetic   &  25.18  &  $\pm$  &  21.28  
								&  5.16  &  $\pm$  & 1.05
								 &  1.46  &  $\pm$  &  0.03  
								&  1.46  &  $\pm$  &  0.02  
								&  0.93  &  $\pm$  &  0.01  
								&  0.69  &  $\pm$  &  0.12  
								&  \textbf{100.00}  &  $\pm$  &  \textbf{0.00} 
				& 166\\
				\cmidrule{2-24}
				& {\tcrppo}   & {25.89}  & $\pm$ &  {29.60} 
                                       &  {5.15}  & $\pm$ &   {2.02}
                                       &  {\textbf{1.55}}  & $\pm$ &   {\textbf{0.19}} 
                                       &  {1.51}  & $\pm$ &   {0.17} 
                                       &  {\textbf{0.97}}  & $\pm$ &   {\textbf{0.03}} 
                                       &  {0.76}  & $\pm$ &   {0.27} 
                                       & {65.23}  & $\pm$ &  {12.47} 
				& {7}\\
				& {\tcrppobuf}  & {\textbf{36.52}}  & $\pm$ &  {\textbf{30.25}} 
                                         &   {5.37}   & $\pm$ &   {1.81}
                                         &   {\textbf{1.55}}   & $\pm$ &   {\textbf{0.17}} 
                                         &   {1.53}   & $\pm$ &   {0.17} 
                                         &   {0.96}   & $\pm$ &   {0.03} 
                                         &   {\textbf{0.83}}   & $\pm$ &   {\textbf{0.21}} 
                                         &   {75.83}  & $\pm$ &   {9.46}
				& {7}\\
				\midrule
				\multirow{9}{*}{\vdjdbP}
				& \randomselect   &  0.08  &  $\pm$  &  0.10  
								&     &  -  &    
								&  1.58  &  $\pm$  &  0.20
								&  \textbf{1.73}  &  $\pm$  &  \textbf{0.28}  
								&  0.94  &  $\pm$  &  0.02  
								&  0.03  &  $\pm$  &  0.03  
								&  95.02  &  $\pm$  &  0.46  
				& \textbf{1}\\
				\cmidrule{2-24}				
				& \mcts   &  0.00  &  $\pm$  &  0.00  
								&     &  -  &     
								 &  0.00  &  $\pm$  &  0.00  
								&  1.35  &  $\pm$  &  0.03   
								&  0.00  &  $\pm$  &  0.00  
								&  0.05  &  $\pm$  &  0.06 
								&  0.46  &  $\pm$  &  0.09
				& 200\\
				& \bpvae   &  0.08  &  $\pm$  &  0.12  
								&    &  -  &     
								  &  1.33  &  $\pm$  &  0.03  
								 &  1.35  &  $\pm$  &  0.01  
								 &  0.95  &  $\pm$  & 0.03 
								 &  0.05  &  $\pm$  &  0.03  
								 &  18.46  &  $\pm$  &  5.13  
				& 9\\
				%
				\cmidrule{2-24}
				& \randommutate (n=5)  &  2.15  &  $\pm$  &  1.82  
								&  \textbf{2.84}  &  $\pm$  & \textbf{1.47}  
								  &  1.48  &  $\pm$  &  0.09  
								 &  1.55  &  $\pm$  &  0.01  
								 &  0.94  &  $\pm$  &  0.02  
								 &  0.16  &  $\pm$  &  0.06  
								 &  99.33  &  $\pm$  &  0.21  
				& 39\\
				& \randommutate (n=10)  &  4.41  &  $\pm$  &  3.72  
								&  3.03  &  $\pm$  & 1.52
								 &  1.48  &  $\pm$  &  0.07  
								&  1.52  &  $\pm$  &  0.02  
								&  0.94  &  $\pm$  &  0.01  
								&  0.24  &  $\pm$  &  0.08  
								&  99.86  &  $\pm$  &  0.08    
				& 77\\
				& \greedy   &  31.81  &  $\pm$  &  17.09  
								&  5.01  &  $\pm$  & 1.12 
								 &  1.47  &  $\pm$  &  0.03  
								&  1.44  &  $\pm$  &  0.02  
								&  0.93  &  $\pm$  &  0.01  
								&  0.64  &  $\pm$  &  0.09  
								&  99.96  &  $\pm$  &  0.06   
				& 71\\
				& \genetic   &  38.57  &  $\pm$  &  20.57  
								&  5.07  &  $\pm$  & 1.07
								 &  1.48  &  $\pm$  &  0.03  
								&  1.47  &  $\pm$  &  0.02  
								&  0.94  &  $\pm$  &  0.01  
								&  0.72  &  $\pm$  &  0.09  
								&  \textbf{100.00}  &  $\pm$  &  \textbf{0.00}  
				& 156\\
				\cmidrule{2-24}
                & {\tcrppo}      & {45.82}  & $\pm$ &  {21.59} 
                               & {5.34}  & $\pm$ &  {1.58}
                               & {1.51}  & $\pm$ &  {0.19} 
                               & {1.51}  & $\pm$ &  {0.19} 
                               & {0.96}  & $\pm$ &  {0.02} 
                               & {0.87}  & $\pm$ &  {0.18} 
                               & {69.99}  & $\pm$ & {11.74}
                & {7} \\
                & {\tcrppobuf}   & {\textbf{58.97}}  & $\pm$ &  {\textbf{29.11}} 
                                      &  {5.20}  & $\pm$ &   {1.69}
                                      &  {\textbf{1.65}}  & $\pm$ &   {\textbf{0.20}} 
                                      &  {1.63}  & $\pm$ &   {0.19} 
                                      &  {\textbf{0.97}}  & $\pm$ &  {\textbf{0.02}} 
                                      &  {\textbf{0.89}}  & $\pm$ &  {\textbf{0.17}}
                                      &  {82.41}  & $\pm$ &   {5.54}
                & {6} \\
				\bottomrule
			\end{tabular}
			\begin{tablenotes}
			\item Columns represent: \qr: the percentage of qualified TCR sequences; 
			        \mutediff:  the edit distance between the original TCR sequences and the mutated qualified TCR sequences; 
			        $\avgistcr(\qTCRs)$/$\avgistcr(\vTCRs)$: the average \istcr scores over the TCRs in set \qTCRs/\vTCRs; 
			        $\avgrecog(\qTCRs)$/$\avgrecog(\vTCRs)$: the average \recog probabilities over the TCRs in set \qTCRs/\vTCRs; 
			        \vr: the percentage of valid TCR sequences.
			        \ncalls: the average number of calls to calculate \reward functions. 
				Values $x\pm y$ represent mean $x$ and standard deviation $y$. 
			        ``-'': not applicable.
			        The ``$n$" value in \randommutate indicates that \randommutate is done $n$ times over each sequence. \par
			\end{tablenotes}
		\end{threeparttable}
	\end{footnotesize}
	\vspace{-5pt}    
\end{sidewaystable}

%% file: tables/peptides.tex
\begin{table}[!h]
  \centering
  \caption{{Peptides in \mcpas and \vdjdb that TCRs are optimized to bind to}}
  \label{tbl:peptides}
  \begin{threeparttable}
    \begin{small}
      \begin{tabular}{
	@{\hspace{2pt}}r@{\hspace{2pt}}    	
	@{\hspace{2pt}}r@{\hspace{2pt}}   
	@{\hspace{2pt}}r@{\hspace{2pt}}  
	@{\hspace{2pt}}r@{\hspace{2pt}}
	}
        \toprule
        \mcpasP & & \vdjdbP \\
        \midrule
        ASNENMETM & & ASNENMETM \\
        ATDALMTGY   & &  ATDALMTGY \\
        HGIRNASFI & & CTPYDINQM \\
        HPKVSSEVHI & & FPRPWLHGL \\
        MEVGWYRSPFSRVVHLYRNGK   & & FRDYVDRFYKTLRAEQASQE \\
        RFYKTLRAEQASQ & & GTSGSPIVNR \\
        SSLENFRAYV & & HGIRNASFI \\
        SSPPMFRV    & & LSLRNPILV \\
        SSYRRPVGI & & NAITNAKII \\
        YSEHPTFTSQY  & & SQLLNAKYL \\
         & & SSLENFRAYV \\
         & & SSPPMFRV \\
         & & SSYRRPVGI \\
         & & STPESANL \\
         & & TTPESANL \\
        \bottomrule
      \end{tabular}
    \end{small}
  \end{threeparttable}
  \vspace{-13pt} 
\end{table}

%% file: tables/setup.tex
\begin{table}[!h]
  \centering
      \caption{Experimental Setup for \tcrppo}
  \label{tbl:setup}

  \begin{threeparttable}
      \begin{tabular}{
	@{\hspace{2pt}}p{0.8\linewidth}@{\hspace{5pt}}
	@{\hspace{2pt}}r@{\hspace{2pt}}      
	}
        \toprule
        description & value\\
        \midrule
        maximum steps $T$       & 8 \\
        threshold of \recog  & 0.9  \\
        threshold of \istcr  & 1.2577  \\
        discount factor $\gamma$  & 0.9 \\
        \midrule
        latent dimension of $\overrightarrow{\hidden}_i/\overleftarrow{\hidden}_i/\hidden^p$ & {256} \\
        hidden dimension of policy \& value networks & {128} \\
        hidden layers of policy \& value networks & 2\\
        \midrule
        number of steps per iteration & 5,120 \\
        batch size for policy update & 64 \\
        number of epochs per iteration & 10 \\
        clip range & 0.2 \\
        learning rate & 3e-4 \\
        coefficient for value function $\alpha_1$ & 0.5 \\
        coefficient for entropy regularization $\alpha_2$ & 0.01  \\
        \bottomrule
      \end{tabular}
  \end{threeparttable}
  \vspace{-10pt} 
\end{table}